\documentstyle[aps,prd]{revtex}
\begin{document}
\hfill{hep-ph/0012140}\par
\hfill{mfute}
\vskip 0.5cm
\centerline{\large\bf Perturbative QCD factorization of
$\pi \gamma^*\to \gamma(\pi)$ and $B\to \gamma(\pi)l\bar \nu$}
\vskip 0.5cm
\centerline{Hsiang-nan Li}
\vskip 0.5cm
\centerline{Department of Physics, National Cheng-Kung University,}\par
\centerline{Tainan, Taiwan 701, Republic of China}
\vskip 0.5cm
\centerline{Physics Division, National Center for Theoretical Sciences,}\par
\centerline{Hsinchu, Taiwan 300, Republic of China}
\vskip 1.0cm
Pacs numbers: 12.38.Bx, 13.40.Gp, 13.25.Hw
\vskip 1.0cm
\centerline{\bf abstract}
\vskip 0.3cm

We prove factorization theorem for the processes $\pi\gamma^*\to\gamma$
and $\pi\gamma^*\to\pi$ to leading twist in the covariant gauge by means
of the Ward identity. Soft divergences cancel and collinear divergences
are grouped into a pion wave function defined by a nonlocal matrix
element. The gauge invariance and universality of the pion wave function
are confirmed. The proof is then extended to the exclusive $B$ meson
decays $B\to\gamma l\bar\nu$ and $B\to\pi l\bar\nu$ in the heavy quark
limit. It is shown that a light-cone $B$ meson wave function, though
absorbing soft dynamics, can be defined in an appropriate frame.
Factorization of the $B\to\pi l\bar\nu$ decay in $k_T$ space, $k_T$ being
parton transverse momenta, is briefly discussed. We comment on the
extraction of the leading-twist pion wave function from experimental
data.

\vskip 2.0cm

\section{INTRODUCTION}

The fundamental concept of perturbative QCD (PQCD) is factorization
theorem, which states that nonperturbative dynamics of a high-energy QCD
process either cancel or can be absorbed into hadron wave functions.
The remaining part, being infrared finite, is calculable in perturbation
theory. A full amplitude is then expressed as the convolution of a hard
amplitude with hadron wave functions. A wave function, because of
its nonperturbative origin, is not calculable. However, PQCD still
possesses a predictive power, since a wave function is universal,
{\it i.e.}, process-independent. With this universality, a wave function
determined by some means, such as QCD sum rules and lattice theory, or
extracted from experimental data, can be employed to make predictions for
other processes involving the same hadron.

Nonperturbative dynamics is reflected by infrared divergences in
radiative corrections. There are two types of infrared divergences, soft
and collinear. Soft divergences come from the region of a loop momentum
$l$, where all its components vanish:
\begin{eqnarray}
l^\mu=(l^+,l^-,l_T)\sim (\lambda,\lambda,\lambda)\;.
\label{sog1}
\end{eqnarray}
The light-cone variables have been adopted, and $\lambda$ is a
small scale. Collinear divergences are associated with a massless quark
of momentum $P=(Q,0,0_T)$, $Q$ being a large scale. In the collienar
region with $l$ parallel to $P$, the components of $l$ behave like
\begin{eqnarray}
l^\mu\sim (Q,\lambda^2/Q,\lambda)\;.
\label{sog2}
\end{eqnarray}
In both regions the invariant mass of the radiated gluon diminishes as
$\lambda^2$, and the corresponding loop integrand may diverge as
$1/\lambda^4$. As the phase space for loop integration vanishes like
$d^4 l\sim \lambda^4$, logarithmic divergences are generated.

Factorization of the above infrared divergences in a QCD process needs to
be performed in momentum, spin, and color spaces. Factorization in
momentum space means that a hard amplitude does not depend on the loop
momentum of a soft or collinear gluon, which has been absorbed into a
hadron wave function. Factorization in spin and color spaces means that
there are separate fermion and color flows between a hard amplitude and a
wave function, respectively. To achieve these, we rely on the
eikonal approximation for loop integrals in leading infrared regions, the
insertion of the Fierz identity to separate fermion flows, and the Ward
identity to sum up diagrams with different color structures. Under the
eikonal approximation, a soft or collinear gluon is detached from the
lines in a hard amplitude and in other wave functions. The Fierz
identity decomposes a full amplitude into contributions characterized by
different twists. The Ward identity is essential for proving
factorization theorem in a nonabelian gauge theory.

In this paper we shall derive the factorization of the scattering
processes $\pi\gamma^*\to\gamma$ and $\pi\gamma^*\to\pi$, which involve
the pion transition form factor and the pion form factor, respectively,
at leading twist using the above techniques. Infrared divergences,
occuring at higher powers of $1/Q^2$, $Q$ being momentum transfer in the
above processes, are neglected. It will be shown that soft divergences
cancel and collinear divergences, factored out of the whole proesses
order by order, are absorbed into a pion wave function, which is defined
by a nonlocal matrix element. The universality of the pion wave function
is equivalent to the universality of the collinear divergences in the two
processes. The definition of the hard amplitude at each order will be
given as a result of the proof. We emphasize that our derivation
is simple, explicitly gauge-invariant, and appropriate for both the
factorizations of the soft and collinear divergences.

The factorization of the process $\pi\gamma^*\to\gamma(\pi)$ has been
proved in \cite{BL}, but in the axial (light-cone) gauge $A^+=0$. In this
gauge factorization automatically holds and the analysis is 
straightforward,
because collinear divergences exist only in reducible diagrams.
Our proof is performed in the covariant gauge, in which collinear
divergences also exist in irreducible diagrams. The colleciton of these
collinear gluons forms a path-ordered integral along the light cone, which
renders a hadron wave function explicitly gauge invariant. The pion wave
function has been constructed from $\gamma^*\gamma\to\pi$ in the
framework of covariant operator product expansion \cite{BFL,MR}. However,
it was not clear how the path-ordered integral is generated \cite{MR}. We
shall demonstrate that it appears as a consequence of the Ward identity.
The factorization of $\pi\gamma^*\to\pi$ has also been proved in \cite{DM}
based on a generalization of the Zimmermann's ''$\Delta$-forest"
prescription \cite{Z}, which involves complicated diagram subtractions.
It will be found that our derivation is simpler.

We then prove factorization theorem for the exclusive $B$ meson decays
$B\to\gamma l\bar\nu$ and $B\to\pi l\bar\nu$, whose topologies are
similar to the scattering processes $\pi\gamma^*\to\gamma$ and
$\pi\gamma^*\to\pi$, respectively. The infrared divergences in
$B\to\gamma l\bar\nu$ and $B\to\pi l\bar\nu$ have been investigated in
\cite{KPY} and \cite{LY1}, respectively. However, a rigorous proof of
factorization theorem has not yet been available. In the heavy quark
limit, terms of $O(\Lambda_{\rm QCD}/M_B)$, $\Lambda_{\rm QCD}$ being the
QCD scale and $M_B$ the $B$ meson mass, are higher-twist and negligible.
It will be shown that in this limit a gauge-invariant $B$ meson wave
function and hard amplitudes can be defined. The $B$ meson wave function
absorbs soft divergences of the above decays, which differ from the
collinear divergences in the pion wave function. However, it is still
possible to construct a light-cone $B$ meosn wave function \cite{GN}, if
an appropriate frame with the photon (pion) moving in the minus or plus
direction is chosen.

We investigate the $O(\alpha_s)$ collinear divergences contained in the
process $\pi\gamma^*\to\gamma$ in Sec.~II, and present the all-order
proof of its factorization theorem by means of the Ward identity in
Sec.~III. The factorization of the process $\pi\gamma^*\to\pi$ is derived
in Sec.~IV. In Sec.~V we prove factorization theorem for the
$B\to\gamma l\bar\nu$ and $B\to\pi l\bar\nu$ decays. Section VI is the
conclusion. The detailed evaluation of the $O(\alpha_s)$ collinear
divergences in $\pi\gamma^*\to\pi$ are supplied in Appendix A. In
Appendix B we comment on the extraction of the leading-twsit pion wave
functions from experimental data of $\pi\gamma^*\to\gamma$ and
$\pi\gamma^*\to\pi$.

\section{$O(\alpha_s)$ FACTORIZATION OF $\pi\gamma^*\to\gamma$}

We start with the factorization of the process $\pi\gamma^*\to\gamma$
at the one-loop level, which will serve as the basis of the all-order
proof. The momentum $P_1$ of the pion and the momentum $P_2$ of the
outgoing on-shell photon are parametrized as
\begin{eqnarray}
& &P_1=(P_1^+,0,{\bf 0}_T)=\frac{Q}{\sqrt{2}}(1,0,{\bf 0}_T)\;,
\nonumber\\
& &P_2=(0,P_2^-,{\bf 0}_T)=\frac{Q}{\sqrt{2}}(0,1,{\bf 0}_T)\;.
\label{mpp}
\end{eqnarray}
Let $\epsilon$ denote the polarization vector of the outgoing photon,
which contains only the transverse components. We consider the kinematic
region with large $Q^2=-q^2$, $q=P_2-P_1$ being the virtual photon
momentum, where PQCD is applicable.

The lowest-order diagrams are displayed in Fig.~1. The lower valence
quark (an anti-quark) in the pion carries the fractional momentum $xP_1$.
Contracting the amplitude in Fig.~1(a) with the leading-twist structure
$\not P_1\gamma_5/\sqrt{2N_c}$ associated with the pion, $N_c=3$ being
the number of colors, we derive
\begin{eqnarray}
H^{(0)}(x)=-ie^2\sqrt{\frac{N_c}{2}}
\frac{tr[\not \epsilon(\not P_2-x\not P_1)\gamma_\mu
\not P_1\gamma_5]}{(P_2-xP_1)^2}
=ie^2\sqrt{\frac{N_c}{2}}
\frac{tr(\not \epsilon\not P_2\gamma_\mu
\not P_1\gamma_5)}{xQ^2}\;.
\label{1a}
\end{eqnarray}
Figure 1(b) leads to the same amplitude, because the pion wave function is
symmetric under the interchange of $x$ and $1-x$. The internal quarks are
regarded as being hard, {\it i.e.}, being off-shell by $O(Q^2)$, since
contributions from the small $x$ region will be suppressed by the pion
wave function, which vanishes like $x$ as $x\to 0$.

We identify the infrared divergences from $O(\alpha_s)$ radiative
corrections \cite{AC,B,KMR} to Fig.~1(a), which are shown in Fig.~2.
Self-energy correction to the internal quark, giving a
next-to-leading-oreder hard amplitude, is not included. Figures 2(a)-2(c)
are the reducible diagrams with the additional gluon attaching the two
valence quarks of the pion. It has been known that soft divergences
cancel among these diagrams. The reason for this cancellation is that
soft gluons, being huge in space-time, do not resolve the color structure
of the pion. Collinear divergences in Figs.~2(a)-2(c) do not cancel,
since the loop momentum flows into the internal quark line in Fig.~2(b),
but not in Figs.~2(a) and 2(c). To absorb the collinear divergences, we
introduce a nonperturbative pion wave function.

The factorization of Figs.~2(a)-2(c) is achieved by meas of the insertion
of the Fierz identity,
\begin{eqnarray}
I_{ij}I_{lk}=\frac{1}{4}I_{ik}I_{lj}
+\frac{1}{4}(\gamma_5)_{ik}(\gamma_5)_{lj}
+\frac{1}{4}(\gamma_\alpha)_{ik}(\gamma^\alpha)_{lj}
+\frac{1}{4}(\gamma_5\gamma_\alpha)_{ik}(\gamma^\alpha\gamma_5)_{lj}
+\frac{1}{8}(\sigma_{\alpha\beta})_{ik}(\sigma^{\alpha\beta})_{lj}\;,
\end{eqnarray}
where $I$ represents the identity matrix, and $\sigma_{\alpha\beta}$ is
defined by $\sigma_{\alpha\beta}\equiv i[\gamma_\alpha,\gamma_\beta]/2$.
Different terms in the above identity lead to contributions of different
twists. Take Fig.~2(b) as an example, whose loop integrand is given by
\begin{eqnarray}
I_{2b}&=& e^2g^2 C_F\sqrt{\frac{N_c}{2}}
tr\Bigg\{\gamma_\nu\frac{x\not P_1-\not l}{(xP_1-l)^2}
\not \epsilon\frac{\not P_2-x\not P_1+\not l}{(P_2-xP_1+l)^2}\gamma_\mu
\nonumber\\
& &\times \frac{(1-x)\not P_1 +\not l}{[(1-x) P_1 +l]^2}\gamma^\nu
\not P_1\gamma_5\Bigg\}\frac{1}{l^2}\;,
\label{2b}
\end{eqnarray}
with $C_F$ being a color factor. Inserting the Fierz identity, we obtain
\begin{eqnarray}
I_{2b}&\approx& ig^2 C_Ftr\Bigg\{\gamma_\nu
\frac{x\not P_1-\not l}{(xP_1-l)^2}
\frac{\gamma_5\gamma_\alpha}{2}
\frac{(1-x)\not P_1 +\not l}{[(1-x) P_1 +l]^2}\gamma^\nu
\frac{\gamma^-\gamma_5}{2}\Bigg\}\frac{1}{l^2}
\nonumber\\
& &\times (-ie^2)\sqrt{\frac{N_c}{2}}
\frac{tr[\not \epsilon(\not P_2-x\not P_1+\not l)\gamma_\mu
\gamma^\alpha\gamma_5)P_1^+}{(P_2-xP_1+l)^2}\;.
\label{2bc}
\end{eqnarray}
Obviously, in the collinear region with the loop momentum $l$ parallel
to $P_1$, only the pseudo-vector structure $\gamma_5\gamma_\alpha$
contributes to the first trace as shown above. Moreover, the matrices
$\gamma_\alpha$ and $\gamma^\alpha$ must be $\gamma_-=\gamma^+$ and
$\gamma^-$, respectively.

Equation (\ref{2bc}), as integrated over $l$, is rewritten as the
convolution of the lowest-order hard amplitude $H^{(0)}(\xi)$ with the
$O(\alpha_s)$ pion wave function $\phi^{(1)}_{2b}$ in the momentum
fraction $\xi=x-l^+/P_1^+$:
\begin{eqnarray}
I_{2b}&\approx& \phi^{(1)}_{2b} H^{(0)}(\xi)\;,
\label{2bco}\\
\phi^{(1)}_{2b}&=& ig^2 C_Ftr\Bigg\{\gamma_\nu
\frac{x\not P_1-\not l}{(xP_1-l)^2}
\frac{\gamma_5\gamma^+}{2}
\frac{(1-x)\not P_1 +\not l}{[(1-x) P_1 +l]^2}\gamma^\nu
\frac{\gamma^-\gamma_5}{2}\Bigg\}\frac{1}{l^2}\;.
\label{p2b}
\end{eqnarray}
$\phi^{(1)}_{2b}$ contains the collinear divergence in Fig.~2(b), because
the integrand in Eq.~(\ref{p2b}) diverges as $1/\lambda^4$. The
dependences on $l^-$ and on $l_T$ in $H^{(0)}$, being subleading
according to Eq.~(\ref{sog2}), have been neglected.

Diagrams with the additional gluon attaching the internal quark,
Figs.~2(d) and 2(e), do not contain soft divergences, because the internal
quark is off-shell. For example, the loop integrand corresponding to
Fig.~2(d) is approximated, in the $l\to 0$ region, by 
\begin{equation}
\frac{1}{(P_2-xP_1+l)^2[(1-x)P_1+l]^2 l^2}\approx
\frac{1}{2(1-x)(P_2-xP_1)^2P_1\cdot l l^2}
\sim O(\lambda^{-3})\;,
\label{2s}
\end{equation}
which is suppressed by the phase space for loop integration
$d^4 l\sim \lambda^4$. Therefore, we consider only the collinear
divergences. The integrand associated with Fig.~2(d) is written as,
\begin{eqnarray}
I_{2d}&=&-e^2g^2 C_F\sqrt{\frac{N_c}{2}}tr\Bigg\{\not \epsilon
\frac{\not P_2-x\not P_1}{(P_2-xP_1)^2}\gamma_\nu
\frac{\not P_2-x\not P_1+\not l}{(P_2-xP_1+l)^2}\gamma_\mu
\nonumber\\
& &\times \frac{(1-x)\not P_1 +\not l}{[(1-x) P_1 +l]^2}\gamma^\nu
\not P_1\gamma_5\Bigg\}\frac{1}{l^2}\;.
\label{2din}
\end{eqnarray}
Since $\not \epsilon$ and $\gamma_\mu$ involve only $\gamma_T$,
we drop $-x\not P_1$ and $-x\not P_1+\not l$ in the internal quark
propagators in the collinear region. Equation (\ref{2din}) is then
simplified into
\begin{eqnarray}
I_{2d}&=&-e^2g^2 C_F\sqrt{\frac{N_c}{2}}\frac{2P_{2\nu}}{(P_2-xP_1+l)^2}
tr\Bigg\{\not \epsilon\frac{\not P_2-x\not P_1}{(P_2-xP_1)^2}
\gamma_\mu\frac{(1-x)\not P_1 +\not l}{[(1-x) P_1 +l]^2}\gamma^\nu
\not P_1\gamma_5\Bigg\}\frac{1}{l^2}\;.
\label{2d}
\end{eqnarray}

For the $l$-dependent hard propagator, we employ the relation
\begin{eqnarray}
\frac{2P_{2\nu}}{(P_2-xP_1+l)^2}\approx\frac{n_\nu}{n\cdot l}
\left[1-\frac{(P_2-xP_1)^2}{(P_2-xP_1+l)^2}\right]\;,
\label{p2id}
\end{eqnarray}
which is an example of the Ward identity. The dimensionless vector
$n=P_2/P_2^-$, representing the direction of an eikonal line, lies along
the outgoing photon momentum. $n_\nu$ is called the eikonal vertex and
$1/n\cdot l$ is called the eikonal propagator. The appearence of the
eikonal line is a consequence of the Ward identity. Inserting the Fierz
identity, we derive the factorization,
\begin{eqnarray}
I_{2d}\approx \phi^{(1)}_{2d}[H^{(0)}(x)-H^{(0)}(\xi)]\;,
\label{6d}
\end{eqnarray}
with the $O(\alpha_s)$ pion wave function
\begin{eqnarray}
\phi^{(1)}_{2d}=-ig^2 C_Ftr\Bigg\{\frac{\gamma_5\gamma^+}{2}
\frac{(1-x)\not P_1 +\not l}{[(1-x) P_1 +l]^2}\gamma^\nu
\frac{\gamma^-\gamma_5}{2}\Bigg\}\frac{1}{l^2}\frac{n_\nu}{n\cdot l}\;.
\label{6dc}
\end{eqnarray}
The collinear contribution from Fig.~2(d) has been split into two terms,
with the first and second terms described by Figs.~3(a) and 3(b),
respectively, where the eikonal propagator is represented by a double
line. In Fig.~3(b) the loop momentum $l$ flows into the internal quark
line, such that the second term is a convolution of $H^{(0)}$
with $\phi_{2d}^{(1)}$ in the momentum fraction $\xi$.

Figure 2(e) gives the loop integrand
\begin{eqnarray}
I_{2e}&=&e^2g^2 C_F\sqrt{\frac{N_c}{2}}tr\Bigg\{\gamma^\nu
\frac{x\not P_1 -\not l}{(x P_1 -l)^2}\not \epsilon
\frac{\not P_2-x\not P_1+\not l}{(P_2-xP_1+l)^2}\gamma_\nu
\nonumber\\
& &\times\frac{\not P_2-x\not P_1}{(P_2-xP_1)^2}\gamma_\mu
\not P_1\gamma_5\Bigg\}\frac{1}{l^2}\;.
\end{eqnarray}
Following the similar procedures, we obtain, in the collinear region,
\begin{eqnarray}
I_{2e}\approx -\phi^{(1)}_{2e}[H^{(0)}(\xi)-H^{(0)}(x)]\;,
\label{6e}
\end{eqnarray}
with the $O(\alpha_s)$ pion wave function
\begin{eqnarray}
\phi^{(1)}_{2e}=ig^2 C_Ftr\Bigg\{\gamma^\nu
\frac{\not P_1 -\not l}{(xP_1 -l)^2}\frac{\gamma_5\gamma^+}{2}
\frac{\gamma^-\gamma_5}{2}\Bigg\}\frac{1}{l^2}\frac{n_\nu}{n\cdot l}\;.
\label{6ec}
\end{eqnarray}
The first and second terms in Eq.~(\ref{6e}) are described by Figs.~3(c)
and 3(d), respectively.

Comparing Eqs.~(\ref{p2b}), (\ref{6dc}) and (\ref{6ec}), the Feynman
rules for the perturbative evaluation of the pion wave function  are
clear: $\phi^{(1)}$ can be written as a nonlocal hadronic matrix element
with the structure $\gamma_5\gamma^+/2$ sandwiched, which comes from the
insertion of the Fierz identity:
\begin{eqnarray}
\phi^{(1)}(x)=\frac{1}{\sqrt{2N_c}P_1^+}
\int \frac{dy^-}{2\pi}e^{ixP_1^+y^-}
\langle 0|{\bar q}(y^-)\frac{\gamma_5\gamma^+}{2}(-ig)
\int_0^{y^-}dzn\cdot A(zn)q(0)|\pi(P_1)\rangle\;.
\label{ld}
\end{eqnarray}
The sum over colors is understood. The integral over $z$ in fact
contains two pieces: For the upper eikonal line in Fig.~3(a), $z$ runs
from 0 to $\infty$. For the lower eikonal line in Fig.~3(b), $z$ runs
from $\infty$ back to $y^-$. It is easy to confirm that the above
definition reproduces all the leading-twist collinear divergences in
Figs.~2(a)-(c) and in Figs.~3(a)-3(d). The light-cone coordinate
$y^-\not =0$ corresponds to the fact that the collinear divergences in
Fig.~2 do not cancel exactly.

At last, the factorization formula for the process $\pi\gamma^*\to\gamma$
is written, up to $O(\alpha_s)$, as
\begin{eqnarray}
(\phi^{(0)}+\phi^{(1)})\otimes(H^{(0)}+H^{(1)})+O(\alpha_s^2)\;,
\end{eqnarray}
with $\phi^{(0)}=1$ and $\otimes$ representing the convolution in the
momentum fraction. The $O(\alpha_s)$ hard amplitude $H^{(1)}$ is defined
by,
\begin{eqnarray}
H^{(1)}\equiv\sum_i \int\frac{d^4 l}{(2\pi)^4}I_i
-\phi^{(1)}\otimes H^{(0)}\;,
\end{eqnarray}
where the subscript $i$ runs from $2a$ to $2e$. Obviously, $H^{(1)}$ is
infrared finite, since all the $(\alpha_s)$ collinear divergences have
been absorbed into the pion wave function $\phi^{(1)}$.

\section{ALL-ORDER PROOF OF FACTORIZATION THEOREM}

In this section we present the all-order proof of leading-twist
factorization theorem for the process $\pi\gamma^*\to \gamma$, and
construct a gauge-invariant pion wave function
defined by
\begin{eqnarray}
\phi(x)=\frac{1}{\sqrt{2N_c}P_1^+}
\int\frac{dy^-}{2\pi}e^{ixP_1^+y^-}
\langle 0|{\bar q}(y^-)\frac{\gamma_5\gamma^+}{2}
P\exp\left[-ig\int_0^{y^-}dzn\cdot A(zn)\right]q(0)|\pi(P_1)
\rangle,
\label{pw}
\end{eqnarray}
as shown in Fig.~3(e). The notation $P$ means the
path ordering. By expanding the quark field ${\bar q}(y^-)$ and the
path-ordered exponential into powers of $y^-$, the above matrix element
can be expressed as a series of covariant derivatives
$(D^+)^n{\bar q}(0)$, implying that Eq.~(\ref{pw}) is gauge invariant.

It has been mentioned in the Introduction that factorization of a QCD
process in momentum, spin, and color spaces requires summations of many
diagrams, especially at higher orders. Hence, the diagram summation
must be handled in an elegant way. For this purpose, we employ the Ward
identity,
\begin{eqnarray}
l_\mu G^\mu(l,k_1,k_2,\cdots, k_n)=0\;,
\label{war}
\end{eqnarray}
where $G^\mu$ represents a physical amplitude with an external gluon
carrying the momentum $l$ and with $n$ external quarks carrying the
momenta $k_1$, $k_2$, $\cdots$, $k_n$. All these external particles are
on mass shell. The Ward identity can be easily derived by means of the
Becchi-Rouet-Stora (BRS) transformation \cite{BRS}.

We prove factorization theorem by induction. The factorization of the
$O(\alpha_s)$ collinear divergences associated with the pion has been
worked out in Sec.~II. Assume that the factorization holds up to
$O(\alpha_s^N)$:
\begin{eqnarray}
G=\phi\otimes H\;,
\label{gh}
\end{eqnarray}
with
\begin{eqnarray}
G=\sum_{i=0}^N G^{(i)}\;,\;\;\;
\phi=\sum_{i=0}^N \phi^{(i)}\;,\;\;\;
H=\sum_{i=0}^N H^{(i)}\;.
\label{ghf}
\end{eqnarray}
$G^{(i)}$ denotes the full diagrams of $O(\alpha_s^i)$ with
$G^{(0)}=H^{(0)}$ in Eq.~(\ref{1a}), the pion wave function $\phi^{(i)}$
is defined by the $O(\alpha_s^i)$ terms in the perturbative expansion of
Eq.~(\ref{pw}), and $H^{(i)}$ is the infrared-finite hard amplitude of
$O(\alpha_s^i)$. We then have the relations
\begin{eqnarray}
G^{(i)}=\sum_{j=0}^i \phi^{(j)}\otimes H^{(i-j)}\;,
\label{ind}
\end{eqnarray}
for $i=0,\cdots, N$, which imply that all collinear divergences in
$G^{(i)}$ have been collected into $\phi^{(j)}$, $j\le i$,
systematically.

Consider a complete set of full diagrams $G^{(N+1)}$ of
$O(\alpha_s^{N+1})$. We look for the gluon, one of whose ends
attaches the outer most vertex on the upper quark line in the  pion. Such
a gluon exists, since $G^{(N+1)}$ are finite-order diagrams. We then
classify $G^{(N+1)}$ into two categories, reducible and irreducible,
according to the attachment of the other end of this gluon. If
the other end attaches the upper or lower quark lines directly, the
diagrams are reducible. The $O(\alpha_s)$ examples are Figs.~2(a)-2(c).
All other diagrams, with the other end attaching inside of the
$O(\alpha_s^N)$ full diagrams $G^{(N)}$, are irreducible. The
$O(\alpha_s)$ examples are Figs.~2(d) and 2(e). The factorization of
reducible diagrams is the same as that of Figs.~2(a)-2(c): we simply
insert the Fierz identity to separate $G^{(N+1)}$ into the convolution of
$G^{(N)}$ with the $O(\alpha_s)$ collinear divergent diagrams, which
contribute to the pion wave function $\phi^{(1)}$. The result is shown in
Fig.~4(a). The explicit expression of $\phi^{(1)}_{2b}$ has been derived
in Eq.~(\ref{p2b}).

We then discuss factorization of irreducible $G^{(N+1)}$. Let $\alpha$ be
the outer most vertex on the upper quark line, and $\beta$ denote the
attachments of the other end of the radiated gluon inside $G^{(N)}$. The
pion possesses the leading-twist spin structure $\not P_1\gamma_5$ from
the Fierz identity. The fermion propagator adjacent to the vertex
$\alpha$ is proportional to $\not P_1$ in the collinear region with the
loop momentum $l$ parallel to $P_1$. Hence, the component $\gamma^+$
in $\gamma^\alpha$, which is located between the spin structure and the
fermion propagator, gives the leading contribution. The vertex
$\beta$ must be dominated by the minus component. With the above
reasoning, we propose the following replacement for the tensor
$g^{\alpha\beta}$ appearing in the propagator of the radiated gluon,
\begin{eqnarray}
g_{\alpha\beta}\to \frac{n_\alpha l_\beta}{n\cdot l}\;.
\label{rep}
\end{eqnarray}
The light-like vector $n_\alpha$ in the minus direction, which was
introduced in Eq.~(\ref{p2id}), selects the plus component of
$\gamma^\alpha$. In the collinear region $l$ lies mainly in the plus
direction, and $l_\beta$ selects the minus component of the vertex $\beta$.
Therefore, Eq.~(\ref{rep}) extracts the leading-twist collinear
divergences from the irreducible $G^{(N+1)}$.

The contraction of $l^\beta$ hints the application of the Ward identity
in Eq.~(\ref{war}) to the case with two external on-shell quarks.
Figure 4(b) contains a complete set of contractions of $l^\beta$, which
are represented by arrows, since the second and third diagrams have been
added. The cuts on the quark lines denote the insertion of the Fierz
identity. The Ward identity, namely, the equation described by
Fig.~4(b), holds. The second diagram gives
\begin{eqnarray}
& &l_\beta\frac{1}{(1-x)\not P_1-\not l}\gamma^\beta \not P_1\gamma_5
\nonumber\\
&=&\frac{1}{(1-x)\not P_1-\not l}[\not l-(1-x)\not P_1 +
(1-x)\not P_1] \not P_1\gamma_5\;,
\nonumber\\
&=&\frac{(1-x)P_1^2\gamma_5}{(1-x)\not P_1-\not l}
-\not P_1\gamma_5\;.
\label{ide}
\end{eqnarray}
The first term in the second expression vanishes because of the
on-shell condition $P_1^2=0$. The second term corresponds to the
$O(\alpha_s^N)$ full diagrams $G^{(N)}$. Similarly, the third diagram
leads to
\begin{eqnarray}
l_\beta\not P_1\gamma_5\gamma^\beta\frac{1}{x\not P_1-\not l}
=-\frac{xP_1^2\gamma_5}{x\not P_1-\not l}-\not P_1\gamma_5\;,
\label{ide2}
\end{eqnarray}
where the first term vanishes and the second term corresponds to
$G^{(N)}$. The factor $n_\alpha/n\cdot l$ from the collinear replacement
in Eq.~(\ref{rep}) is exactly the Feynman rule associated with the eikonal
line in the direction of $n$. Equations (\ref{ide}) and (\ref{ide2}) 
imply that in the collinear region the irreducible $G^{(N+1)}$ are
factorized as shown in Fig.~4(c). Obviously, the factorization of the
irreducible diagrams with the gluon emitted from the outer most vertex
on the lower quark line exists. We conclude that the irreducible
$G^{(N+1)}$ can be factorized into the convolution of $G^{(N)}$ with
$\phi^{(1)}$ described by Figs.~3(a)-3(d), in which the radiated gluon
attaches the eikonal lines.

We derive the collinear factorization of $G^{(N+1)}$,
\begin{eqnarray}
G^{(N+1)}\approx \phi^{(1)}\otimes G^{(N)}\;,
\label{inc}
\end{eqnarray}
where $\phi^{(1)}$ contains both the reducible gluons in Fig.~4(a) and
the irreducible gluons in Figs.~3(a)-3(d). The remaining part
$F^{(N+1)}$, defined via
\begin{eqnarray}
G^{(N+1)}= \phi^{(1)}\otimes G^{(N)}+F^{(N+1)}\;,
\label{wi1}
\end{eqnarray}
is infrared finite, {\it i.e.}, free from the collinear divergence
associated with the pion. The above procedures are also applicable to
the $O(\alpha_s^{N+1})$ pion wave function $\phi^{(N+1)}$, which is
defined by the perturbative expansion of Eq.~(\ref{pw}). We have
\begin{eqnarray}
\phi^{(N+1)}= \phi^{(1)}\otimes \phi^{(N)}+{\bar F}^{(N+1)}\;,
\label{wi2}
\end{eqnarray}
with the infrared-finite function ${\bar F}^{(N+1)}$.

Employing Eqs.~(\ref{ind}), (\ref{wi1}), and (\ref{wi2}), we write
\begin{eqnarray}
G^{(N+1)}&=&\phi^{(1)}\otimes\sum_{j=0}^N \phi^{(j)}\otimes H^{(N-j)}
+F^{(N+1)}
\nonumber\\
&=&\sum_{j=0}^N [\phi^{(j+1)}-{\bar F}^{(j+1)}]\otimes H^{(N-j)}
+F^{(N+1)}
\nonumber\\
&=&\sum_{j=1}^{N+1} \phi^{(j)}\otimes H^{(N+1-j)}+H^{(N+1)}\;,
\label{fac1}
\end{eqnarray}
with the $O(\alpha_s^{N+1})$ hard amplitude,
\begin{eqnarray}
H^{(N+1)}\equiv F^{(N+1)}-\sum_{j=1}^N {\bar F}^{(j+1)}\otimes H^{(N-j)}\;.
\label{hao}
\end{eqnarray}
Obviously, the function $H^{(N+1)}$ does not contain any collinear
divergence, since both $F$ and $\bar F$ are infrared finite. Because of
$\phi^{(0)}=1$, Eq.~(\ref{fac1}) becomes
\begin{eqnarray}
G^{(N+1)}=\sum_{j=0}^{N+1} \phi^{(j)}\otimes H^{(N+1-j)}\;.
\label{fac2}
\end{eqnarray}
The above expression indicates that all collinear divergences in the full
diagrams of $\pi\gamma^*\to\gamma$ can be factored into the definition
of the pion wave function in Eq.~(\ref{pw}) order by order, and that the
remaining hard amplitude is infrared finite. Allowing $N$ to approach
infinity, we prove factorization theorem for the process
$\pi\gamma^*\to\gamma$ to all orders.

At last, we prove by induction that soft divergences do not exist in the
process $\pi\gamma^*\to\gamma$ to leading twist. The $O(\alpha_s)$ soft
cancellation has been explained in Sec.~II. Assume that the
$O(\alpha_s^N)$ full diagrams $G^{(N)}$ do not contain any soft
divergence (though they contain collinear divergences). Consider the
$O(\alpha_s^{N+1})$ full diagrams $G^{(N+1)}$. Similarly, we look for the
gluon radiated from the outer most vertex on the upper quark line, and
classify the diagrams into the reducible and irreducible types. For
reducible diagrams, we express $G^{(N+1)}$ as the convolution of
$G^{(N)}$ with the $O(\alpha_s)$ infrared divergent diagrams as shown in
Fig.~4(a). It is easy to confirm that the derivation in
Eqs.~(\ref{2b})-(\ref{2bco}) still applies to the soft region with the loop
momentum $l\to 0$. The soft divergences in the $O(\alpha_s)$
diagrams cancel in the same way as in Figs.~2(a)-2(c). The remaining
$G^{(N)}$ are free from soft divergences as assumed above. Hence,
the reducible $G^{(N+1)}$ have no soft divergences.

We then consider irreducible $G^{(N+1)}$. The diagrams $G^{(N)}$, without
soft divergences, are dominated either by hard or by collinear dynamics.
In the hard region of $G^{(N)}$, internal particle propagators behave
like $1/Q^2$. The attachment of a soft gluon, producing one more hard
propagator, does not introduce soft divergences. The reason for this
absence of soft divergences is the same as in Eq.~(\ref{2s}). In the
collinear region of $G^{(N)}$, where momenta parallel to $P_1$ dominate,
we employ the eikonal approximation for the small loop momentum $l$,
\begin{eqnarray}
\frac{\not P_1+\not l}{(P_1+l)^2}\gamma_\alpha\not P_1
\approx \frac{P_{1\alpha}}{P_1\cdot l}\not P_1\;.
\label{eik}
\end{eqnarray}
The contraction of the numerator $P_{1\alpha}$ to the outer most vertex
$\gamma^\alpha$, which is mainly $\gamma^+$, leads to a vanishing
contribution. Therefore, the irreducible $G^{(N+1)}$ do not contain soft
divergences either. Extending $N$ to infinity, we prove that the process
$\pi\gamma^*\to\gamma$ is free from soft divergences.

\section{FACTORIZATION OF $\pi\gamma^*\to\pi$}

We investigate infrared divergences in the process $\pi\gamma^*\to\pi$,
and discuss only the factorization of the initial-state pion wave
function. The discussion of the final-state wave function is the same.
Assume that the incoming and outgoing pions carry the momenta $P_1$ and
$P_2$, respectively, which are defined by Eq.~(\ref{mpp}), and that the
momentum fraction $x_1$ ($x_2$) is associated with the lower quark line
in the incoming (outgoing) pion. The above kinematic variables have been
indicated in the lowest-order diagrams in Fig.~5. Similarly, we consider
the region with large momentum transfer $Q^2=-q^2$, $q=P_2-P_1$ being the
virtual photon momentum.

Contracting the four-quark amplitudes in Fig.~5 with the leading-twist
spin structures $\not P_1\gamma_5/\sqrt{2N_c}$ and
$\gamma_5\not P_2/\sqrt{2N_c}$ from the initial and final states,
respectively, we derive
\begin{eqnarray}
& &H_{5a}^{(0)}(x_1,x_2)=\frac{i}{2}eg^2 C_F
\frac{tr[\gamma^\nu\gamma_5\not P_2\gamma_\nu(\not P_2-x_1\not P_1)
\gamma_\mu\not P_1\gamma_5]}{(P_2-x_1P_1)^2(x_1P_1-x_2P_2)^2}
=-\frac{4eg^2C_F}{x_1x_2 Q^2}iP_{1\mu}\;,
\\
& &H_{5b}^{(0)}(x_1,x_2)=\frac{i}{2}eg^2 C_F
\frac{tr[\gamma^\nu\gamma_5\not P_2\gamma_\mu
(\not P_1-x_2\not P_2)\gamma_\nu\not P_1\gamma_5]}
{(P_1-x_2P_2)^2(x_1P_1-x_2P_2)^2}
=-\frac{4eg^2C_F}{x_1x_2 Q^2}iP_{2\mu}\;.
\end{eqnarray}
The other two lowest-order diagrams, where the virtual photon attaches
the lower quark lines, lead to the same expressions but with different
electric charge $e$. Since contributions from soft partons
($x_1,x_2\to 0$) are suppressed by the pion wave functions, the exchanged
gluons in Fig.~5, off-shell by $O(Q^2)$, are regarded as being hard.

\subsection{$O(\alpha_s)$ Factorization}

We first identify infrared divergences from $O(\alpha_s)$ radiative
corrections \cite{MN} to Fig.~5(a). The diagrams in Fig.~6 contain
potential infrared divergences associated with the incoming pion. We do
not consider self-energy corrections to the inernal lines, since they,
without infrared divergences, give only a next-to-leading-order hard
amplitude. The treatment of Figs.~6(a)-6(c) is the same as that of
Figs.~2(a)-2(c): soft divergences cancel and collinear divergences are
factored into the pion wave function $\phi^{(1)}$ by inserting the Fierz
identity.

Diagrams with the additional gluon attaching the internal lines, such as
Figs.~6(d)-6(g), do not generate soft divergences, because the internal
lines are off-shell. We concentrate only on collinear
divergences. Following the detailed calculation in Appendix A, the
collinear divergences in Figs.~6(d) and 6(e) are given by
\begin{eqnarray}
I_{6d}&\approx& \frac{N_c}{2C_F}\phi_{2d}^{(1)}
[H_{5a}^{(0)}(x_1,x_2)-H_{5a}^{(0)}(\xi_1,x_2)]\;,
\label{2d3}\\
I_{6e}&\approx& \frac{N_c}{2C_F}\phi_{2e}^{(1)}
[H_{5a}^{(0)}(x_1,x_2)-H_{5a}^{(0)}(\xi_1,x_2)]\;,
\label{2e3}
\end{eqnarray}
with $\xi_1=x_1-l^+/P_1^+$. The dependences on $l^-$ and $l_T$ in
$H_{5a}^{(0)}$ have been neglected, since they are subleading in the
collinear region. Collinear divergences from $l$ parallel to $P_1$ vanish
in Figs.~6(f) and 6(g). We have separated the integrands $I_{6d}$ and
$I_{6e}$ into two terms, similar to those in Eqs.~(\ref{6d}) and
(\ref{6e}). However, the color factors are different because of the
different color flows in Figs.~2(d) and 6(d). We shall explain how to fix
the color factors after calculating Figs.~6(h)-6(k).

The collinear factorization of the irreducible corrections in
Figs.~6(h)-6(k) relies on the eikonal approximation. The results are
written as
\begin{eqnarray}
I_{6h}&\approx&-\frac{1}{2C_FN_c}\phi_{2d}^{(1)}
H_{5a}^{(0)}(x_1,x_2)\;,
\label{2hp}\\
I_{6i}&\approx&\frac{1}{2C_FN_c}\phi_{2d}^{(1)}
H_{5a}^{(0)}(\xi_1,x_2)\;,
\label{2ia}\\
I_{6j}&\approx&\frac{1}{2C_FN_c}\phi_{2e}^{(1)}
H_{5a}^{(0)}(\xi_1,x_2)\;,
\label{2jb}\\
I_{6k}&\approx&-\frac{1}{2C_FN_c}\phi_{2e}^{(1)}
H_{5a}^{(0)}(x_1,x_2)\;.
\label{2ka}
\end{eqnarray}
To discuss soft divergences, we simply allow $l^+$ to appraoch zero,
{\it i.e.}, $\xi_1\to x_1$. It is easy to find that soft divergences
cancel between Figs.~6(h) and 6(i) and between Figs.~6(j) and 6(k).
Hence, irreducible corrections are free from soft divergences.

Combining Eq.~(\ref{2hp}) and Eq.~(\ref{2ia}) with the first and second
terms in Eq.~(\ref{2d3}), respectively, and combining Eq.~(\ref{2jb})
and Eq.~(\ref{2ka}) with the second and first terms in Eq.~(\ref{2e3}),
respectively, we obtain the correct color factors:
\begin{eqnarray}
I_{6(d)-6(k)}&\approx&\phi_{2d}^{(1)}
[H_{5a}^{(0)}(x_1,x_2)-H_{5a}^{(0)}(\xi_1,x_2)]
\nonumber\\
& &+\phi_{2e}^{(1)}[H_{5a}^{(0)}(x_1,x_2)-H_{5a}^{(0)}(\xi_1,x_2)]\;,
\label{3d}
\end{eqnarray}
which are also described by Figs.~3(a)-3(d). It is observed from the
above expression that the collinear divergences in the summation of
Figs.~6(d)-6(k) are identical to those in the summation of Figs.~2(d)
and 2(e), consistent with the universality of hadron wave functions.

We then investigate infrared divergences from radiative corrections to
Fig.~5(b). The diagrams in Fig.~7 contain potential infrared divergences
associated with the incoming pion. The factorization of Figs.~7(a)-7(c)
is the same as that of Figs.~6(a)-6(c): They are absorbed into
the leading-twist pion wave function $\phi^{(1)}$ straightforwardly.
The contributions from Figs.~7(d) and 7(e) are the same as those
from Figs.~6(d) and 6(e).

For Figs.~7(f), we consider only the collinear divergences from $l$
parallel to $P_1$,
\begin{eqnarray}
I_{7f}&\approx&\frac{ig^2}{N_c} \frac{n\cdot [(1-x_1) P_1 +l]}
{[(1-x_1) P_1 +l]^2l^2n\cdot(P_1+l)}H_{5b}^{(0)}(x_1,x_2)\;.
\label{4fa}
\end{eqnarray}
The result differs form that of Fig.~6(f), which vanishes. Note that the
collinear divergence in Eq.~(\ref{4fa}) is not in the correct eikonal
form. To show that it does not cause any trouble, we consider the
collinear factorization of Fig.~7(h),
\begin{eqnarray}
I_{7h}&\approx&\frac{ig^2}{N_c} \frac{n\cdot [(1-x_1) P_1 +l]}
{[(1-x_1) P_1 +l]^2l^2n\cdot(P_1+l)}\frac{n\cdot P_1}{n\cdot l}
H_{5b}^{(0)}(x_1,x_2)\;.
\label{4ha}
\end{eqnarray}
Combining Eqs.~(\ref{4fa}) and (\ref{4ha}), we have the simpler expression
\begin{eqnarray}
I_{7f+7h}&\approx&-\frac{1}{2C_FN_c}\phi_{2d}^{(1)}
H_{5b}^{(0)}(x_1,x_2)\;,
\label{4fh}
\end{eqnarray}
whose collinear divergence is exactly the same as that of Fig.~6(h) in
Eq.~(\ref{2hp}). At last, combining the collinear divergences in
Eq.~(\ref{4fh}) and those from Fig.~7(d), we obtain the result described
by Fig.~3(a).

A similar analysis shows that the collinear divergences in the sum of
Figs.~7(g) and 7(j) are the same as in Fig.~6(j). Combined with the
divergences from Fig.~7(e), we obtain the result described by Fig.~3(c).
The results for Figs.~7(i) and 7(k) are the same as for Figs.~6(i)
and 6(k), respectively, and their collinear factorizations, combined with
Figs.~7(d) and 7(e), are described by Figs.~3(b) and 3(d), respectively.
We conclude that the $O(\alpha_s)$ collinear divergences in Figs.~6 and 7
are identical to those in Fig.~2, and can be formulated into the
leading-twist pion wave function $\phi^{(1)}$ defined in Eq.~(\ref{ld}).

\subsection{All-order Factorization}

The analysis of the $O(\alpha_s)$ collinear divergences in
$\pi\gamma^*\to \pi$ indicates that factorization of a QCD process
requires summations of many diagrams. For example, the summation of
Figs.~7(f) and 7(h) removes the dependence of the hard amplitude on the
loop momentum $l$. The summations of Figs.~6(d) and 6(h) and of
Figs.~7(d), 7(f) and 7(h) give the correct color factors. To prove
factorization theorem to all orders, we deal
with these summations by means of the Ward identity. It is trivial to
generalize the proof for the process $\pi\gamma^*\to\gamma$ to
$\pi\gamma^*\to\pi$. In the present case the amplitude $G^\mu$ in
Eq.~(\ref{war}) contains four external on-shell quarks. We simply
interpret the function $H$ as the part that does not involve the
collinear divergences associated with the incoming pion. Then repeat the
procedures and further factorize $H$ into the the convolution of a
hard amplitude with the wave function associated with the outgoing pion.

The proof of the absence of soft divergences in the process
$\pi\gamma^*\to \pi$ is subtler. It has been shown that the $O(\alpha_s)$
corrections do not produce soft divergences. Assume that the soft
cancellation exists up to $O(\alpha_s^N)$. For reducible
$O(\alpha_s^{N+1})$ full diagrams $G^{(N+1)}$, we insert the Fierz
identity to factor out the $O(\alpha_s)$ infrared divergent diagrams,
which involve the gluons emitted from the outer most vertices on the
upper and lower quark lines. This insertion works for both the initial-
and final-state pions. The soft divergences in the $O(\alpha_s)$ diagrams
cancel in the same way as in Figs.~2(a)-2(c). The remaining
$O(\alpha_s^N)$ full diagrams $G^{(N)}$ have no soft divergences as assumed
above. Hence, the reducible $G^{(N+1)}$ do not contain soft
divergences.

For irreducible $G^{(N+1)}$, we consider the gluon emitted from the outer
most vertices on the upper and lower quark lines in the incoming pion.
Since $G^{(N)}$ are free from soft divergences, they are dominated either
by hard or by collinear dynamics. In the hard region and in the collinear
region with momenta parallel to $P_1$, the soft divergences vanish for the
same reason as for the process $\pi\gamma^*\to\gamma$. In the collinear
region with momenta parallel to $P_2$, we adopt the eikonal
approximation,
\begin{eqnarray}
\frac{\not P_2+\not l}{(P_2+l)^2}\gamma_\alpha\not P_2
\approx \frac{n_\alpha}{n\cdot l}\not P_2\;.
\label{eik2}
\end{eqnarray}
It is easy to observe that the above approximation can be achieved by the
replacement in Eq.~(\ref{rep}).

We propose to employ Eq.~(\ref{rep}) to extract the soft divergences
in the irreducible $G^{(N+1)}$. In the hard region and in the collinear
region with momenta parallel to $P_1$, the replacement in Eq.~(\ref{rep})
simply modifies subleading contributions, that are free from soft
divergences. In the collinear region with momenta parallel to $P_2$, the
replacement extracts the correct soft divergences in the irreducible
$G^{(N+1)}$. Hence, Eq.~(\ref{rep}) always works for a leading-twist
analysis of soft divergences. The derivation from Eq.~(\ref{ide}) to
Eq.~(\ref{inc}) holds, and the irreducible $G^{(N+1)}$ are factorized
into the convolution of $G^{(N)}$ with Figs.~3(a)-3(d). Since $G^{(N)}$
have no soft divergences, we allow the loop momentum $l$ flowing inside
$G^{(N)}$ to approach zero. It is then obvious that the soft
divergences cancel between Figs.~3(a) and 3(b) and bwteen Figs.~3(c) and
3(d). That is, the irreducible $G^{(N+1)}$, like the reducible
$G^{(N+1)}$, are also free from soft divergences. The same
argument applies to the gluons emitted from the outgoing pion side.
At last, extending $N$ to infinity, we prove the absence of soft
divergences in the process $\pi\gamma^*\to \pi$.

We conclude from the discussions in Secs.~III and IV that the
leading-twist infrared structures in the processes $\pi\gamma^*\to\pi$
and $\pi\gamma^*\to\gamma$ are identical, and that the pion wave function,
defined by Eq.~(\ref{pw}), is universal.

\section{FACTORIZATION OF $B$ MESON DECAYS}

In this section we present the all-order proof of factorization theorem
for the exclusive $B$ meson decays $B\to \gamma l\bar\nu$ and
$B\to\pi l\bar\nu$ in the heavy quark limit. It will be shown that 
the proof for the pion transition form factor in Sec.~III
can be generalized to $B$ meson decays, if terms suppressed by powers of
$1/M_B$ are neglected. The momentum $P_1$ of the $B$ meson and the
momentum $P_2$ of the outgoing on-shell photon (pion) are parametrized as
\begin{eqnarray}
P_1=\frac{M_B}{\sqrt{2}}(1,1,{\bf 0}_T)\;,\;\;\;
P_2=\frac{M_B}{\sqrt{2}}(0,\eta,{\bf 0}_T)\;,
\label{bmpp}
\end{eqnarray}
where $\eta$ denotes the energy fraction carried by the photon (pion).
Assume that the light spectator quark in the $B$ meson carries the
momentum $k$, and that $\epsilon$ denotes the polarization vector of the
photon. We consider the kinematic region with small $q^2$,
$q=P_1-P_2$ being the lepton pair momentum, {\it i.e.}, with large
$\eta$, where PQCD is applicable. In the heavy quark limit the mass
difference between the $B$ meson and the $b$ quark, $\bar\Lambda=M_B-m_b$,
is a small scale, which will appear as higher-twist terms proportional
to ${\bar\Lambda}/M_B$. The four components of the spectator quark
momentum $k$ are of the same order as $\bar\Lambda$.

The lowest-order diagrams for the $B\to\gamma l\bar\nu$ decay and for the
$B\to\pi l\bar\nu$ decay are displayed in Fig.~1 and Fig.~5, respectively,
but with the upper quark line replaced by a $b$ quark line and with the
vertex $\times$ representing a weak decay vertex. It is easy to observe
that Figs.~1(a) and 1(b) scale like
\begin{eqnarray}
H_{1a}^{(0)}&\propto&
\frac{1}{(P_2-k)^2}\sim\frac{1}{{\bar\Lambda} M_B}\;,
\label{b1a}\\
H_{1b}^{(0)}&\propto&
\frac{1}{(q-k)^2-M_B^2}\sim \frac{1}{M_B^2}\;,
\label{b1b}
\end{eqnarray}
indicating that Fig.~1(b) is power-suppressed. In the following analysis 
we shall concentrate only on Fig.~1(a). There are two leading-twist spin 
structures for the $B$ meson,
\begin{eqnarray}
\frac{(\not P_1+M_B){\not{\bar n}}\gamma_5}{2\sqrt{N_c}}\;,\;\;\;
\frac{(\not P_1+M_B){\not n}\gamma_5}{2\sqrt{N_c}}\;,
\label{sbs}
\end{eqnarray}
with the dimensionless vecor $\bar n=(1,0,0_T)$,
each of which is associated with a $B$ meson wave function. For the 
radiative decay $B\to\gamma l\bar\nu$, only the first structure
contributes at the leading-twist.

Contracting the amplitudes in Fig.~1(a) with 
$(\not P_1+M_B){\not{\bar n}}\gamma_5/(2\sqrt{N_c})$, we derive
\begin{eqnarray}
H_{1a}^{(0)}(x)=-e\frac{\sqrt{N_c}}{2}
\frac{tr[\not \epsilon\not P_2\gamma_\mu(1-\gamma_5)
\not n]}{x\eta M_B}\;,
\end{eqnarray}
where $x=k^+/P_1^+$ is the momentum fraction. Since $P_2$ has been chosen 
in the minus direction, only the plus component $k^+$ of $k$ is relevant.
We have dropped the higher-twist terms proportional to $k$ in the
numerators and $\bar\Lambda$ by assuming $m_b\approx M_B$.
It is then possible to integrate out the $k^-$
and $k_T$ dependences in the $B$ meson wave function $\psi_+(k)$,
leading to the light-cone $B$ meson wave function,
\begin{eqnarray}
\phi_+(x)=\int dk^-d^2k_T \psi_+(k)\;.
\end{eqnarray}
It will be shown that a light-cone $B$ meson wave function can
always be defined, if an appropriate frame is chosen, even though the
four components of $k$ are of the same order.


\subsection{$O(\alpha_s)$ Factorization of $B\to\gamma l\bar\nu$}

We start with the one-loop diagrams in Fig.~2 for the
$B\to\gamma l\bar\nu$ decay, discussing the factorization of their
infrared divergences. The factorization of Figs.~2(a)-2(c) requires the
insertion of the Fierz identity. Take Fig.~2(b) as an example (the
analysis of Figs.~2(a) and 2(c) is trivial). The loop integrand is
written as
\begin{eqnarray}
I_{2b}&\approx& ig^2 C_Ftr\Bigg\{\gamma_\nu\frac{\not k-\not l}{(k-l)^2}
\frac{\gamma_5\gamma^\alpha}{2}
\frac{\not P_1-\not k +\not l+M_B}{(P_1-k +l)^2-M_B^2}\gamma^\nu
\frac{(\not v+I){\not{\bar n}}\gamma_5}{2}\Bigg\}\frac{1}{l^2}
\nonumber\\
& &\times e\frac{\sqrt{N_c}}{2}
\frac{tr[\not \epsilon(\not P_2-\not k+\not l)\gamma_\mu(1-\gamma_5)
M_B\gamma_\alpha\gamma_5]}{(P_2-k+l)^2}\;,
\label{b2bc}
\end{eqnarray}
where we have kept only the leading-twist structure
$\gamma_5\gamma^\alpha$ in the Fierz identity, and introduced the
dimensionless vector $v=P_1/M_B$. The other structures, such as 
$\gamma_5$ and $\gamma_5\sigma^{\alpha\beta}$, do not
contribute, because the corresponding hard amplitude vanishes:
\begin{eqnarray}
e\frac{\sqrt{N_c}}{2}
\frac{tr[\not \epsilon(\not P_2-\not k+\not l)\gamma_\mu(1-\gamma_5)
M_B\gamma_5]}{(P_2-k+l)^2}=0\;.
\end{eqnarray}

In the heavy quark limit we have the eikonal approximation,
\begin{eqnarray}
\frac{\not P_1-\not k +\not l+M_B}{(P_1-k +l)^2-M_B^2}\gamma^\nu
(\not v+I)
\approx \frac{v^\nu}{v\cdot (l-k)}(\not v+I)\;,
\label{hq}
\end{eqnarray}
which is in fact equivalent to the heavy quark expansion. Obviously, 
only the component $\gamma_+=\gamma^-$ of $\gamma_\alpha$ 
contributes to the hard amplitude at leading twist, requiring the spin 
structure $\gamma_5\gamma^\alpha$ to be $\gamma_5\gamma^+$ in the first 
trace for the $B$ meson wave function. Equation (\ref{b2bc}) becomes
\begin{eqnarray}
I_{2b}&\approx& (\phi_+^{(1)})_{2b}H_{1a}^{(0)}(\xi)\;,
\end{eqnarray}
with $\xi=x-l^+/P_1^+$ and the $O(\alpha_s)$ $B$ meson wave function,
\begin{eqnarray}
(\phi_+^{(1)})_{2b}=
ig^2 C_Ftr\Bigg\{\gamma_\nu\frac{\not k-\not l}{(k-l)^2}
\frac{\gamma_5\gamma^+}{2}
\frac{(\not v+I){\not{\bar n}}\gamma_5}{2}\Bigg\}\frac{1}{l^2}
\frac{v^\nu}{v\cdot (l-k)}\;.
\label{b2b2}
\end{eqnarray}
Figure 2(b) has been expressed as the convolution of the hard
amplitude $H_{1a}^{(0)}$ with the $O(\alpha_s)$ infrared divergent
diagrams, which contribute to $\phi_+^{(1)}$, in the momentum fraction
$\xi$.

Equation (\ref{b2b2}) is simplified into
\begin{eqnarray}
(\phi_+^{(1)})_{2b}=-ig^2 C_F\frac{2v^-(l^+-k^+)}{(k-l)^2l^2
v\cdot (l-k)}\;.
\end{eqnarray}
Performing the contour integration over, say, $l^-$, we observe that the
integral is singular only when the component $l^+$ is of 
$O(\bar\Lambda)$. This observation implies
that the infrared divergence associated with the $B$ meson is of the soft
type. This dynamics differs from the collinear type of divergences
associated with the pion in the process $\pi\gamma^*\to\gamma$. It is easy
to understand that the soft divergences in Figs.~2(a)-2(c) do not cancel
in $B$ meson decays \cite{LY1}: The light spectator quark, carrying a
small amount of momenta, forms a color cloud around the $b$ quark. This
cloud is also huge in space-time, such that soft gluons resolve the
color structure of the $B$ meson.

Diagrams with the additional gluon attaching the internal quark in
Figs.~2(d) and 2(e) also contain soft divergences, since the
internal quark is off-shell only by $O(\bar\Lambda M_B)$. Note that the
internal quark in the process $\pi\gamma^*\to\gamma$ is off-shell by
$O(Q^2)$. There is no collinear divergence from $l$ parallel to $P_2$,
because the internal quark propagator remains the scaling law
$1/\bar\Lambda M_B$, instead of $1/\bar\Lambda^2$. The integrand
associated with Fig.~2(d) is written as,
\begin{eqnarray}
I_{2d}&=&-ieg^2 C_F\frac{\sqrt{N_c}}{2}tr\Bigg\{\not \epsilon
\frac{\not P_2-\not k}{(P_2-k)^2}\gamma_\nu
\frac{\not P_2-\not k+\not l}{(P_2-k+l)^2}\gamma_\mu(1-\gamma_5)
\nonumber\\
& &\times \frac{\not P_1-\not k +\not l+M_B}{(P_1-k +l)^2-M_B^2}
\gamma^\nu(\not P_1+M_B){\not{\bar n}}\gamma_5\Bigg\}\frac{1}{l^2}\;.
\end{eqnarray}
Inserting the Fierz identity, we have,
\begin{eqnarray}
I_{2d}&\approx& -ig^2 C_F\frac{2P_{2\nu}}{(P_2-k+l)^2}
tr\Bigg\{\frac{\gamma_5\gamma^\alpha}{2}
\frac{\not P_1-\not k +\not l+M_B}{(P_1-k +l)^2-M_B^2}\gamma^\nu
\frac{(\not v+I){\not{\bar n}}\gamma_5}{2}\Bigg\}\frac{1}{l^2}
\nonumber\\
& &\times e\frac{\sqrt{N_c}}{2}
\frac{tr[\not \epsilon\not P_2\gamma_\mu(1-\gamma_5)
M_B\gamma_\alpha\gamma_5]}{(P_2-k)^2}\;.
\label{2df}
\end{eqnarray}
Obviously, the first trace is proportional to ${\bar n}^\alpha$ in the 
heavy quark limit.

We employ the relation similar to Eq.~(\ref{p2id}),
\begin{eqnarray}
\frac{2P_{2\nu}}{(P_2-k+l)^2}\approx
\frac{n_\nu}{n\cdot l}\left[1-\frac{(P_2-k)^2}{(P_2-k+l)^2}\right]\;,
\label{p2i}
\end{eqnarray}
where the terms $(k-l)^2\sim O({\bar\Lambda}^2)$ and
$k^2\sim O({\bar\Lambda}^2)$ have been neglected. The eikonal line in the
direction of $n$ appears. Equation (\ref{2df}) reduces to
\begin{eqnarray}
I_{2d}\approx (\phi_+^{(1)})_{2d}[H_{1a}^{(0)}(x)-H_{1a}^{(0)}(\xi)]\;,
\end{eqnarray}
with the $O(\alpha_s)$ $B$ meson wave function,
\begin{eqnarray}
(\phi_+^{(1)})_{2d}=
-ig^2 C_F\frac{n_\nu}{n\cdot l}
tr\Bigg\{\frac{\gamma_5\gamma^+}{2}
\frac{(\not v+I){\not{\bar n}}\gamma_5}{2}\Bigg\}\frac{1}{l^2}
\frac{v^\nu}{v\cdot (l-k)}\;.
\label{b6d}
\end{eqnarray}
The above expression implies that the infrared divergences in irreducible
diagrams can also be collected by the eikonal line along the light cone.
This is attributed to the choice of the frame, in which the photon moves
in the minus direction. The first and second terms in Eq.~(\ref{b6d}) are
described by Figs.~8(a) and 8(b), respectively, where the eikonal lines
in the directions of $v$ and of $n$ have been indicated. In Fig.~8(b) the
loop momentum $l$ flows into the internal quark line, such that the
second term appears as a convolution of $H_{1a}^{(0)}$
with $(\phi_+^{(1)})_{2d}$ in the momentum fraction $\xi$.

Figure 2(e) gives the loop integrand
\begin{eqnarray}
I_{2e}&=&ieg^2 C_F\frac{\sqrt{N_c}}{2}tr\Bigg\{\gamma^\nu
\frac{\not k -\not l}{(k -l)^2}\not \epsilon
\frac{\not P_2-\not k+\not l}{(P_2-k+l)^2}\gamma_\nu
\nonumber\\
& &\times\frac{\not P_2-\not k}{(P_2-k)^2}\gamma_\mu(1-\gamma_5)
(\not P_1+M_B){\not{\bar n}}\gamma_5\Bigg\}\frac{1}{l^2}\;.
\end{eqnarray}
Inserting the Fierz identity, we have,
\begin{eqnarray}
I_{2e}&\approx& ig^2 C_F\frac{2P_{2\nu}}{(P_2-k+l)^2}
tr\Bigg\{\gamma^\nu\frac{\not k -\not l}{(k -l)^2}
\frac{\gamma_5\gamma^\alpha}{2}
\frac{(\not v+I){\not{\bar n}}\gamma_5}{2}\Bigg\}\frac{1}{l^2}
\nonumber\\
& &\times e\frac{\sqrt{N_c}}{2}
\frac{tr[\not \epsilon\not P_2\gamma_\mu(1-\gamma_5)
M_B\gamma_\alpha\gamma_5]}{(P_2-k)^2}\;.
\label{2ef}
\end{eqnarray}
Similarly, the component $\gamma_+=\gamma^-$ of $\gamma_\alpha$ in the 
second trace is selected. Employing Eq.~(\ref{p2i}), the above 
expression is rewritten as
\begin{eqnarray}
I_{2e}\approx -(\phi_+^{(1)})_{2e}[H_{1a}^{(0)}(\xi)-H_{1a}^{(0)}(x)]\;,
\end{eqnarray}
with the $O(\alpha_s)$ $B$ meson wave function
\begin{eqnarray}
(\phi_+^{(1)})_{2e}=ig^2 C_F\frac{n_\nu}{n\cdot l}
tr\Bigg\{\gamma^\nu\frac{\not k -\not l}{(k -l)^2}
\frac{\gamma_5\gamma^+}{2}
\frac{(\not v+I){\not{\bar n}}\gamma_5}{2}\Bigg\}\frac{1}{l^2}\;.
\label{b2e}
\end{eqnarray}
We have split Fig.~2(e) into two terms, whose infrared divergent factors
are described by Figs.~8(c) and 8(d), respectively

Comparing Eqs.~(\ref{b2b2}), (\ref{b6d}) and (\ref{b2e}), the Feynman
rules for the perturbative evaluation of the $B$ meson wave function are
clear: $\phi_B^{(1)}$ can be written as a nonlocal hadronic matrix element
with the structure $\gamma_5\gamma^+/2$ sandwiched:
\begin{eqnarray}
\phi_+^{(1)}(x)=\frac{1}{\sqrt{2N_c}M_B}
\int \frac{dy^-}{2\pi}e^{ixP_1^+y^-}
\langle 0|{\bar q}(y^-)\frac{\gamma_5\gamma^+}{2}(-ig)
\int_0^{y^-}dzn\cdot A(zn)b_v(0)|B(P_1)\rangle\;,
\label{b1}
\end{eqnarray}
where the rescaled $b$ quark field,
\begin{eqnarray}
b_v(y)=\exp(iM_B v\cdot y)\frac{\not v+I}{2}b(y)\;,
\end{eqnarray}
has been introduced. The Feynman rules associated with $b_v$ are those
for an eikonal line in the direction of $v$ given in Eq.~(\ref{hq}). The
above definition reproduces the contributions from Figs.~2(a)-2(c) and
from Figs.~8(a)-8(d), if it is evaluated perturbatively.

The factorization formula for the $B\to\gamma l\bar\nu$ decay
is written, up to $O(\alpha_s)$, as
\begin{eqnarray}
(\phi_+^{(0)}+\phi_+^{(1)})\otimes(H_{1a}^{(0)}+H_{1a}^{(1)})
+O(\alpha_s^2)\;,
\end{eqnarray}
with $\phi_+^{(0)}=1$ and $\otimes$ representing the convolution in the
momentum fraction. The $O(\alpha_s)$ hard amplitude $H_{1a}^{(1)}$ is
defined by,
\begin{eqnarray}
H_{1a}^{(1)}\equiv\sum_i \int\frac{d^4 l}{(2\pi)^4}I_i
-\phi_+^{(1)} \otimes H_{1a}^{(0)}\;,
\end{eqnarray}
where the subscript $i$ runs from $2a$ to $2e$. Obviously, $H_{1a}^{(1)}$
is infrared finite, since all the $(\alpha_s)$ soft divergences have
been absorbed into the $B$ meson wave function $\phi_+^{(1)}$.

\subsection{All-order Factorization of $B\to\gamma l\bar\nu$}

We prove leading-twist factorization theorem for the
$B\to \gamma l\bar\nu$ decay to all orders, and construct a
gauge-invariant light-cone $B$ meson wave function defined by
\begin{eqnarray}
\phi_+(x)&=&\frac{1}{\sqrt{2N_c}M_B}
\int\frac{dy^-}{2\pi}e^{ixP_1^+y^-}
\langle 0|{\bar q}(y^-)\frac{\gamma_5\gamma^+}{2}
P\exp\left[-ig\int_0^{y^-}dzn\cdot A(zn)\right]b_v(0)|B(P_1)
\rangle\;,
\nonumber\\
& &
\label{bw}
\end{eqnarray}
as shown in Fig.~8(e). The eikonal lines in the
directions of $v$ and of $n$ have been indicated. By expanding the
light quark field ${\bar q}(y^-)$ and the path-ordered exponential into
powers of $y^-$, the nonlocal matrix element can be expressed as a series
of covariant derivatives $(D^+)^n{\bar q}(0)$, implying that
Eq.~(\ref{bw}) is gauge invariant.

We present the proof by induction. The factorization of the $O(\alpha_s)$
infrared divergences associated with the $B$ meson has been worked out.
Assume that the factorization of the infrared divergences holds up to
$O(\alpha_s^N)$, that is, we have Eqs.~(\ref{gh})-(\ref{ind}) for the
$B\to\gamma l\bar\nu$ decay. Consider a complete set of
$O(\alpha_s^{N+1})$ full diagrams $G^{(N+1)}$. We look for the 
gluon, one of whose ends attaches the outer most vertex on the $b$ quark
line. We classify $G^{(N+1)}$ into the reducible and irreducible types
as in Sec.~III. The factorization of the reducible $G^{(N+1)}$ is the
same as that of Figs.~2(a)-2(c): Following Eqs.~(\ref{b2bc})-(\ref{b2b2}),
we insert the Fierz identity to separate the reducible $G^{(N+1)}$ into
the convolution of $G^{(N)}$ with the $O(\alpha_s)$ infrared divergent
diagrams, which contribute to the $B$ meson wave function $\phi_B^{(1)}$.
The result is similar to that shown in Fig.~4(a).

For the factorization of the irreducible diagrams, we rely on the Ward
identity in Eq.~(\ref{war}). In this case the amplitude $G^\mu$ contains
two on-shell external quarks, one of which is the heavy $b$ quark.
As hinted by the $O(\alpha_s)$ analysis, the soft divergence,
associated with the gluon radiated by the outer most vertex on
the $b$ quark line, can be collected by the eikonal approximation,
{\it i.e.}, by the replacement in Eq.~(\ref{rep}). The reason is as
follows. In the heavy quark limit the full diagrams of the
$B\to\gamma l\bar\nu$ decay are dominated by the momentum flow along the
photon momentum $P_2$ in the minus direction. Strickly speaking, they are
dominated by $P_2-k$. Hence, the vertex $\beta$ inside $G^{(N)}$ the
radiated gluon attaches is mainly minus, and the vertex $\alpha$ on the
$b$ quark line is mainly plus. Apparently, the tensor
$n_\alpha l_\beta/n\cdot l$ extracts the correct leading contribution.
Therefore, we have the equation described by Fig.~4(b).

The second and third diagrams in Fig.~4(b) give
\begin{eqnarray}
& &l_\beta\frac{1}{\not P_1-\not k-\not l-M_B}\gamma^\beta
(\not v+I){\not{\bar n}}\gamma_5
\nonumber\\
&=&-\frac{1}{\not P_1-\not k-\not l-M_B}\not k
(\not v+I){\not{\bar n}}\gamma_5-(\not v+I){\not{\bar n}}\gamma_5\;,
\label{bide}\\
& &l_\beta (\not v +I){\not{\bar n}}
\gamma_5\gamma^\beta\frac{1}{\not k-\not l}
=(\not v +I){\not{\bar n}}\gamma_5\not k\frac{1}{\not k-\not l}
-(\not v+I){\not{\bar n}}\gamma_5\;,
\label{bide2}
\end{eqnarray}
respectively. After assigning the quark propagators on the right-hand
sides of the above expressions into the corresponding loop integrals, the
first terms are proportional to $k\sim O(\bar\Lambda)$. They are
suppressed by $O(\bar\Lambda/M_B)$ compared to the second terms, which
correspond to $G^{(N)}$.
Neglecting the higher-twist terms, the irreducible $G^{(N+1)}$ are
factorized into the convolution of $G^{(N)}$ with the $O(\alpha_s)$ $B$
meson wave function, in which the radiated gluon attaches the eikonal
lines as in Figs.~8(a) and 8(b). The factorization of the irreducible
$G^{(N+1)}$, with the gluon emitted from the outer most vertex on the
light spectator quark line, is similar. The resultant $O(\alpha_s)$
infrared divergent diagrams for the $B$ meson wave function are those
in Figs.~8(c) and 8(d).

Combining the factorizations of the reducible and irreducible $G^{(N+1)}$,
we arrive at Eq.~(\ref{wi1}). The factorization in Eq.~(\ref{wi2}) for
the $B$ meson wave function $\phi_+^{(N+1)}$ also exists. Following the
steps in Eqs.~(\ref{fac1})-(\ref{fac2}), we complete the all-order proof
of leading-twist factorization theorem for the $B\to\gamma l\bar\nu$
decay. The definition of the hard amplitude at each order is the same as
in Eq.~(\ref{hao}). Since the full diagrams are dominated by momenta
along $P_2$, only the plus component $k^+$ of $k$ is relevant in the hard
amplitude. This is the reason we can integrate the $B$ meson wave function
over $k^-$ and $k_{T}$, obtaining the light-cone $B$ meson wave
function in Eq.~(\ref{bw}).

\subsection{Factorization of $B\to\pi l\bar\nu$}

The factorization of the pion wave function in Secs.~II, III, IV, and the
factorization of the $B$ meson wave function in the $B\to\gamma l\bar\nu$
decay can be applied to the $B\to\pi l\bar\nu$ decay straightforwardly.
Here we simply explain some points of the proof, and neglect the details. 
The lowest-order diagrams are shown in Fig.~5. Both the spin structures 
in Eq.~(\ref{sbs}) contribute to Fig.~5(b), but only the first one 
contributes to Fig.~5(a). Since $P_2$ has been chosen in the minus 
direction, only the plus component $k^+$ of $k$ is relevant. It is then 
possible to define a light-cone $B$ meson wave function for the 
$B\to\pi l\nu$ decay. We first identify infrared divergences from 
$O(\alpha_s)$ radiative corrections to Fig.~5(a), which are displayed in 
Fig.~6, since their analysis is similar to that of the 
$B\to\gamma l\bar\nu$ decay. Note that Fig.~6 is complete only for the 
construction of the $B$ meson wave function. We do not consider 
self-energy corrections to the internal lines, which give only 
next-to-leading-order hard amplitudes.

The construction of the pion wave function is
basically the same as that for the process $\pi\gamma^*\to \pi$. That is,
the infrared divergences associated with the outgoing pion is of the
collinear type. Soft divergences cancel by pairs, because soft 
gluons do not interact with the color singlet pion. For example, soft
divergences cancel between Fig.~6(h) and 6(i) and between 6(j) and 6(k).
The collinear gluons are still collected by the eikonal lines along the
light cone, even though they may attach the heavy $b$ quark. The reason
is that when a loop momentum $l$ is parallel to the pion momentum $P_2$
in the minus direction, only the component $\gamma^+$ of the vertex on
the $b$ quark line and $P_1^+$ of the $B$ meson momentum are selected.
Consequently, we derive the definition for the pion wave function, which
is identical to Eq.~(\ref{pw}). This conclusion is consistent with the
universality of hadron wave functions.

We then concentrate on the $B$ meson wave function. The treatment of
the reducible diagrams in Figs.~6(a)-6(c) is exactly the same as of
Figs.~2(a)-2(c) for the $B\to\gamma l\bar\nu$ decay. We insert the Fierz
identity to separate these diagrams into the convolution of Fig.~5(a)
with the $O(\alpha_s)$ soft divergent diagrams, which contribute
to the $B$ meson wave function $\phi_+^{(1)}$. The investigation of the
irreducible diagrams in Figs.~6(d)-6(e) hints that the soft divergences
can be collected by the eikonal lines along the direction of $n$, a
conclusion similar to that for Figs.~2(d) and 2(e). Hence, the
replacement in Eq.~(\ref{rep}) for the loop gluon extracts the soft
divergences of the irreducible diagrams in Figs.~6(d)-6(k). The Ward
identity applies, and the sum of these irreducible diagrams is factorized
into the convolution of Fig.~5(a) with part of $\phi_+^{(1)}$ described
by Figs.~8(a)-8(d). Following the procedures in Sec.~III, we prove 
leading-twist factorization theorem for Fig.~5(a) in the 
$B\to\pi l\bar\nu$ decay. The $B$ meson wave functions $\phi_+$ 
constructed from the $B\to\gamma l\bar\nu$ and $B\to\pi l\bar\nu$ decays
are identical.

The discussion of the $O(\alpha_s)$ radiative corrections to Fig.~5(b) 
shown in Fig.~7 is similar, though more complicated. The soft 
approximation, the Fierz insertion, and the Ward identity apply. 
Because of the existence of the heavy quark propagator, the spin 
structures $\gamma_5\gamma^\alpha$, $\gamma_5$ and 
$\gamma_5\sigma^{\alpha\beta}$ in the Fierz identity contribute.
In this case two leading-twist $B$ meson wave functions can be 
constructed \cite{GN}. Besides $\phi_+$ defined in Eq.~(\ref{bw}), we 
derive the additional light-cone $B$ meson wave function associated with
the structure $(\not P_1+M_B){\not n}\gamma_5/(2\sqrt{N_c})$, 
\begin{eqnarray}
\phi_-(x)&=&\frac{1}{\sqrt{2N_c}M_B}
\int\frac{dy^-}{2\pi}e^{ixP_1^+y^-}
\langle 0|{\bar q}(y^-)\frac{\gamma_5\gamma^-}{2}
P\exp\left[-ig\int_0^{y^-}dzn\cdot A(zn)\right]b_v(0)|B(P_1)
\rangle\;.
\nonumber\\
& &
\label{bwm}
\end{eqnarray}
In summary, the nonlocal hadronic matrix element for the $B$ 
meson is expressed as 
\begin{eqnarray}
\int \frac{dy^-}{2\pi}e^{ixP_1^+y^-}
\langle 0|{\bar q}_{\gamma}(y^-)b_{\beta}(0)|B(P_1)\rangle
=-\left\{\frac{(\not P_1+M_B)}{\sqrt{2N_c}}
\left[\frac{{\not{\bar n}}}{\sqrt{2}}\phi_{+}(x)
+\frac{\not n}{\sqrt{2}}\phi_{-}(x)\right]\gamma_5\right\}_{\gamma\beta}.
\label{bwp}
\end{eqnarray}
This result is exactly the same as that obtained in \cite{GN}
at leading twist.

A remark is in order. The hard amplitude from Fig.~5(b), proportional to
$1/(x_1 x_2^2)$, $x_2$ being the momentum fraction associated with the
pion, develops an infrared singularity, if the pion wave function
vanishes like $x_2$ as $x_2\to 0$ \cite{ASY}. Even though we have
proved leading-twist factorization theorem for the $B\to\pi l\bar\nu$
decay, its practical application is questionable. It has been shown that
the inclusion of parton transverse momenta $k_T$ smears the singularities
from the end-point region of momentum fractions \cite{LY1}. When $k_T$
is included, the dependence on the transverse loop momentum $l_T$,
being of the same order as $k_T$, is not negligible in the hard
amplitude. This complexity can be resolved by Fourier transforming the
factorization formula from the $k_T$ space into the $b$ space, where $b$
is the impact parameter conjugate to $k_T$ \cite{LS,BS}. The $l_T$
dependence then appears in the factor $\exp(-i{\bf l}_T\cdot {\bf b})$,
and decouples from the hard ampltude. This factor can be assigned into
the corresponding loop integral, which contributes to the definition of
the meson wave functions. We claim that the factorization of the
$B\to\pi l\bar\nu$ decay must be performed in the impact parameter
space \cite{L2}.

\section{CONCLUSION}

In this paper we have investigated the infrared divergences in the
processes $\pi\gamma^*\to \gamma$ and $\pi\gamma^*\to\pi$. We summarize
their comparision below. There are no soft divergences, since they are
either absent or cancel among sets of diagrams. In the collinear region
with $l$ parallel to $P_1$, Figs.~2(a)-2(c) are identical to
Figs.~6(a)-6(c) [Figs.~7(a)-7(c)]. Figures 2(d) and 2(e) are identical to
the combination of Figs.~6(d)-6(k) [Figs.~7(d)-7(k)]. That is, the
collinear structures are the same at the leading twist, consistent with
the concept of universality of hadron wave functions in PQCD
factorization theorem. However, due to the potential significant
subleading contributions at low energies, the extraction of the
leading-twist pion wave function from the $\pi\gamma^*\to\gamma(\pi)$
data suffers ambiguity. For details, refer to Appendix B.

We have presented an all-order proof of leading-twist factorization
theorem for the processes $\pi\gamma^*\to\gamma$ and $\pi\gamma^*\to\pi$,
and for the decays $B\to\gamma l\bar\nu$ and $B\to\pi l\bar\nu$ by means
of the Ward identity. The small scales, such as the light spectator quark
momentum $k$ and the $B$ meson and $b$ quark mass difference
$\bar\Lambda$, are neglected in the heavy quark limit, such that the Ward
identity holds. Our proof is simple compared to that in \cite{DM}, and
explicitly gauge invariant, compared to that in \cite{BL}. We have
constructed the gauge-invariant pion and $B$ meson wave functions, and
confirmed their universality. The path-ordered integral appears as a
consequence of the Ward idenity. Note the difference between the
definitions for the pion and $B$ meson wave functions. The former,
collecting the collinear divergences, depends on the structure
$\gamma_5\gamma^+$. The latter, collecting the soft divergences, depends
on the structure $\gamma_5\not v$. We emphasize that it is possible to
define a light-cone $B$ meson wave function, if an appropriate frame is
chosen, in which the photon (pion) moves in the minus or plus direction.
This is also the reason we can extract the infrared divergences in
$\pi\gamma^*\to\gamma(\pi)$ and $B\to\gamma(\pi) l\bar\nu$ using the same
replacement in Eq.~(\ref{rep}).

The leading-twist factorization of the $B\to\pi l\bar\nu$ decay can be
proved straightforwardly, following the procedures presented in
Secs.~II-V. However, for a practical application, the parton transverse
momenta $k_T$ must be included in order to smear the end-point
singularities in the hard amplitude. We shall derive factorization
theorem including these additional degrees of freedom elsewhere
\cite{L2}. Our proof will be generalized to nonleptonic $B$ meson decays,
such as $B\to\pi\pi$. This factorization is more complicated, since
nonleptonic decays involve three characteristic scales: the $W$ boson
mass $M_W$, $M_B$, and small scales of $O(\bar\Lambda)$, such as $k_T$.

\vskip 0.5cm

I thank S. Brodsky, S. Dorokhov, T. Morozumi, M. Nagashima, D. Pirjol,
A.I. Sanda, and G. Sterman for useful discussions. This work was
supported in part by the National Science Council of R.O.C. under the
Grant No. NSC-89-2112-M-006-033, by Grant-in Aid for Special Project
Research (Physics of CP Violation) from the Ministry of Education,
Science and Culture and by Hiroshima University, Japan.

\vskip 0.5cm

\section*{APPENDIX A: $O(\alpha_s)$ COLLINEAR CORRECTIONS}

In this Appendix we supply the details of the derivation of the collinear
divergences in the process $\pi\gamma^*\to\pi$. The loop integrand from
Fig.~6(d) is written as
\begin{eqnarray}
I_{6d}&=&\frac{-ieg^4}{2N_c}tr\Bigg\{\gamma^\lambda
\gamma_5\not P_2\gamma^\beta
\frac{\not P_2-x_1\not P_1+\not l}{(P_2-x_1P_1+l)^2}\gamma_\mu
\frac{(1-x_1)\not P_1 +\not l}{[(1-x_1) P_1 +l]^2}
\nonumber\\
& &\times \gamma^\alpha\not P_1\gamma_5\Bigg\}
\frac{tr(T^aT^bT^c)\Gamma^{abc}_{\alpha\beta\lambda}}
{l^2(x_1P_1-x_2P_2-l)^2(x_1P_1-x_2P_2)^2}\;,
\label{2d0}
\end{eqnarray}
with $N_c$ being the number of colors and the triple-gluon vertex,
\begin{eqnarray}
\Gamma^{abc}_{\alpha\beta\lambda}&=&-f^{abc}
[g_{\alpha\beta}(2l-x_1P_1+x_2P_2)_\lambda
+g_{\beta\lambda}(2x_1P_1-2x_2P_2-l)_\alpha
\nonumber\\
& &+g_{\lambda\alpha}(x_2P_2-x_1P_1-l)_\beta]\;,
\label{tg}
\end{eqnarray}
$f^{abc}$ being a antisymmetric tensor. In the collinear region with
$l$ parallel to $P_1$, only the term proportional to $g_{\beta\lambda}$ is
important, since the terms proportional to $g_{\lambda\alpha}$ and
$g_{\alpha\beta}$, giving
\begin{eqnarray}
& &[(1-x_1)\not P_1 +\not l]
\gamma^\alpha\not P_1\gamma_5\gamma^\lambda g_{\lambda\alpha}
=2P_1\cdot[(1-x_1)P_1 +l]\gamma_5 \sim O(\lambda^2)\;,
\nonumber\\
& &\gamma^\alpha\not P_1\gamma_5(2\not l-x_1\not P_1+x_2\not P_2)
\gamma_5\not P_2\gamma^\beta g_{\alpha\beta}
=2P_1\cdot(2l-x_1P_1)\not P_2 \sim O(\lambda^2) \not P_2\;,
\end{eqnarray}
do not produce collinear divergences. Considering the second term in
Eq.~(\ref{tg}), Eq.~(\ref{2d0}) exhibits the collinear divergence
\begin{eqnarray}
I_{6d}\approx \frac{-3ig^2P_2\cdot [(1-x)P_1 +l]}
{[(1-x) P_1 +l]^2l^2[P_2\cdot (l-xP_1)]}H_{5a}^{(0)}\;,
\label{2d2f}
\end{eqnarray}
where we have employed the identities,
\begin{eqnarray}
tr(T^aT^bT^c)=\frac{1}{4}(d^{abc}+if^{abc})\;,\;\;\;
d^{abc}f^{abc}=0\;,\;\;\;f^{abc}f^{abc}=24\;,
\end{eqnarray}
$d^{abc}$ being a symmetric tensor, and the approximation for the
denominator
\begin{eqnarray}
(x_1P_1-x_2P_2-l)^2=2x_2P_2\cdot(l-x_1P_1)\;.
\end{eqnarray}
It is found that the infrared divergent piece
of the radiative correction has been completely factored out.
The remaining part, denoted by the lowest-order hard amplitude
$H_{5a}^{(0)}$, does not depend on the loop momentum $l$ at all.

Figure 6(e) gives the loop integrand
\begin{eqnarray}
I_{6e}&=&\frac{ieg^4}{2N_c}tr\Bigg\{\gamma^\alpha
\frac{x_1\not P_1 -\not l}{(x_1 P_1 -l)^2}\gamma^\lambda
\gamma_5\not P_2\gamma^\beta
\frac{\not P_2-x_1\not P_1}{(P_2-x_1P_1)^2}
\nonumber\\
& &\times\gamma_\mu\not P_1\gamma_5\Bigg\}
\frac{tr(T^aT^bT^c)\Gamma^{abc}_{\alpha\beta\lambda}}
{l^2(x_1P_1-x_2P_2-l)^2(x_1P_1-x_2P_2)^2}\;,
\label{2e0}
\end{eqnarray}
with the triple-gluon vertex,
\begin{eqnarray}
\Gamma^{abc}_{\alpha\beta\lambda}&=&-f^{abc}
[g_{\alpha\beta}(l-x_1P_1+x_2P_2)_\lambda
+g_{\beta\lambda}(2x_1P_1-2x_2P_2+l)_\alpha
\nonumber\\
& &+g_{\lambda\alpha}(x_2P_2-x_1P_1-2l)_\beta]\;.
\end{eqnarray}
Similarly, in the region with $l$ parallel to $P_1$ we have
\begin{eqnarray}
I_{6e}\approx \frac{-3ig^2}{(x_1P_1 -l)^2l^2}H_{5a}^{(0)}\;.
\label{2e2f}
\end{eqnarray}

The integrand associated with Fig.~6(f) is written as,
\begin{eqnarray}
I_{6f}&=&\frac{eg^4 C_F^2}{2}tr\Bigg\{
\gamma_\alpha \gamma_5\not P_2\gamma^\alpha
\frac{\not P_2-x_1\not P_1}{(P_2-x_1P_1)^2}\gamma_\nu
\frac{\not P_2-x_1\not P_1+\not l}{(P_2-x_1P_1+l)^2}\gamma_\mu
\nonumber\\
& &\times \frac{(1-x_1)\not P_1 +\not l}{[(1-x_1) P_1 +l]^2}\gamma^\nu
\not P_1\gamma_5\Bigg\}\frac{1}{l^2(x_1P_1-x_2P_2)^2}\;,
\end{eqnarray}
which is simplified into
\begin{eqnarray}
I_{6f}=-2eg^4 C_F^2
\frac{tr[\gamma_\mu(\not P_2+\not l)\not P_1\not l]}
{[(1-x_1) P_1 +l]^2(P_2-x_1P_1+l)^2 l^2(x_1P_1-x_2P_2)^2}\;.
\label{2f}
\end{eqnarray}
The trace in the above expression is proportional to the vanishing
factor $P_1\cdot l$, which suppresses the divergence from the denominator
$[(1-x_1) P_1 +l]^2$. Figure 6(g) does not contain collinear divergences
for the same reason.

The loop integrand associated with Fig.~6(h) is writen as
\begin{eqnarray}
I_{6h}&=&-\frac{eg^4 C_F}{4N_c}tr\Bigg\{\gamma_\alpha\gamma_5\not P_2
\gamma_\nu\frac{(1-x_2)\not P_2+\not l}{[(1-x_2)P_2+l]^2}
\gamma^\alpha\frac{\not P_2-x_1\not P_1 +\not l}{(P_2-x_1 P_1 +l)^2}
\gamma_\mu
\nonumber\\
& &\times \frac{(1-x_1)\not P_1 +\not l}{[(1-x_1)P_1+l]^2}
\gamma^\nu\not P_1\gamma_5
\Bigg\}\frac{1}{l^2(x_1P_1-x_2P_2)^2}\;,
\label{2h}
\end{eqnarray}
which is simplified into
\begin{eqnarray}
I_{6h}&=&-\frac{2eg^4 C_F}{N_c}
\frac{P_1\cdot [(1-x_2)P_2+l]}{[(1-x_2)P_2+l]^2}
tr\Bigg\{\not P_2
\frac{\not P_2-x_1\not P_1 +\not l}{(P_2-x_1 P_1 +l)^2}
\gamma_\mu
\nonumber\\
& &\times \frac{(1-x_1)\not P_1 +\not l}{[(1-x_1)P_1+l]^2}
\Bigg\}\frac{1}{l^2(x_1P_1-x_2P_2)^2}\;.
\end{eqnarray}
In the region with $l$ parallel to $P_1$, we have the approximation,
\begin{eqnarray}
I_{6h}\approx\frac{ig^2}{N_c}
\frac{P_2\cdot [(1-x_1)P_1 +l]}{[(1-x_1)P_1+l]^2l^2P_2\cdot l}H_{4a}^{(0)}
=\frac{ig^2}{N_c}\frac{n\cdot [(1-x_1)P_1 +l]}
{[(1-x_1)P_1+l]^2l^2n\cdot l}H_{5a}^{(0)}\;,
\label{2hpf}
\end{eqnarray}
where we have adopted the approximation for the quark propagator
$[(1-x_2)P_2 +l]\approx 2(1-x_2)P_2\cdot l$, and dropped all the terms
proportional to $P_1\cdot l$.

The loop integrand corresponding to Fig.~6(i) is written as
\begin{eqnarray}
I_{6i}&=&\frac{eg^4 C_F}{4N_c}tr\Bigg\{\gamma_\alpha
\frac{x_2\not P_2+\not l}{(x_2P_2+l)^2}\gamma_\nu\gamma_5\not P_2
\gamma^\alpha\frac{\not P_2-x_1\not P_1 +\not l}{(P_2-x_1 P_1 +l)^2}
\gamma_\mu
\nonumber\\
& &\times \frac{(1-x_1)\not P_1 +\not l}{[(1-x_1)P_1+l]^2}
\gamma^\nu\not P_1\gamma_5
\Bigg\}\frac{1}{l^2(x_1P_1-x_2P_2-l)^2}\;,
\end{eqnarray}
whose sign is opposite to that of Eq.~(\ref{2h}) due to the anti-quark
propagator. The above expression is simplified into
\begin{eqnarray}
I_{6i}&=&\frac{2eg^4 C_F}{N_c}
tr\Bigg\{\frac{x_2\not P_2+\not l}{(x_2P_2+l)^2}
\frac{\not P_2-x_1\not P_1 +\not l}{(P_2-x_1 P_1 +l)^2}
\gamma_\mu
\nonumber\\
& &\times \frac{(1-x_1)\not P_1 +\not l}{[(1-x_1)P_1+l]^2}
\Bigg\}\frac{P_1\cdot P_2}{l^2(x_1P_1-x_2P_2-l)^2}\;.
\end{eqnarray}
It is easy to derive the collinear approximation,
\begin{eqnarray}
I_{6i}\approx \frac{-ig^2}{N_c}
\frac{n\cdot [(1-x_1)P_1 +l]}{[(1-x_1)P_1+l]^2l^2 n\cdot l}
\frac{(x_1P_1-x_2P_2)^2}{(x_1P_1-x_2P_2-l)^2}H_{5a}^{(0)}\;.
\label{2iaf}
\end{eqnarray}
The similar procedures apply to Figs.~6(j) and 6(k).

The integrand associated with Fig.~7(f) is written as
\begin{eqnarray}
I_{7f}&=&-\frac{eg^4 C_F}{4N_c}tr\Bigg\{\gamma_\alpha
\gamma_5\not P_2\gamma_\mu
\frac{\not P_1-x_2\not P_2}{(P_1-x_2P_2)^2}\gamma_\nu
\frac{\not P_1-x_2\not P_2+\not l}{(P_1-x_2P_2+l)^2}\gamma^\alpha
\nonumber\\
& &\times \frac{(1-x_1)\not P_1+\not l}{[(1-x_1) P_1 +l]^2}\gamma^\nu
\not P_1\gamma_5\Bigg\}\frac{1}{l^2(x_1P_1-x_2P_2)^2}\;,
\end{eqnarray}
which is simplified into
\begin{eqnarray}
I_{7f}&=&-\frac{2eg^4 C_F}{N_c} \frac{P_1\cdot(l-x_2P_2)}{(P_1-x_2P_2+l)^2}
tr\Bigg\{\not P_2\gamma_\mu
\frac{\not P_1-x_2\not P_2}{(P_1-x_2P_2)^2}
\nonumber\\
& &\times \frac{(1-x_1)\not P_1+\not l}{[(1-x_1) P_1 +l]^2}
\Bigg\}\frac{1}{l^2(x_1P_1-x_2P_2)^2}\;.
\end{eqnarray}
In the collinear region we have the approximation,
\begin{eqnarray}
I_{7f}\approx \frac{ig^2}{N_c} \frac{P_2\cdot [(1-x_1) P_1 +l]}
{[(1-x_1) P_1 +l]^2l^2P_2\cdot(P_1+l)}H_{5b}^{(0)}\;.
\label{4fai}
\end{eqnarray}

The integrand corresponding to Fig.~7(h) is written as
\begin{eqnarray}
I_{7h}&=&-\frac{eg^4 C_F}{4N_c}tr\Bigg\{\gamma_\alpha
\gamma_5\not P_2\gamma_\nu
\frac{(1-x_2)\not P_2+\not l}{[(1-x_2)P_2+l]^2}\gamma_\mu
\frac{\not P_1-x_2\not P_2+\not l}{(P_1-x_2P_2+l)^2}\gamma^\alpha
\nonumber\\
& &\times \frac{(1-x_1)\not P_1+\not l}{[(1-x_1) P_1 +l]^2}\gamma^\nu
\not P_1\gamma_5\Bigg\}\frac{1}{l^2(x_1P_1-x_2P_2)^2}\;,
\end{eqnarray}
which is simplified into
\begin{eqnarray}
I_{7h}&=&-\frac{2eg^4 C_F}{N_c}
\frac{P_2\cdot[(1-x_1)P_1+l]}{[(1-x_1) P_1 +l]^2}
tr\Bigg\{\frac{(1-x_2)\not P_2+\not l}{[(1-x_2)P_2+l]^2}
\gamma_\mu
\nonumber\\
& &\times \frac{\not P_1-x_2\not P_2+\not l}{(P_1-x_2P_2+l)^2}
\not P_1\Bigg\}\frac{1}{l^2(x_1P_1-x_2P_2)^2}\;.
\end{eqnarray}
We have the collinear approximation,
\begin{eqnarray}
I_{7h}\approx \frac{ig^2}{N_c} \frac{P_2\cdot [(1-x_1) P_1 +l]}
{[(1-x_1) P_1 +l]^2l^2P_2\cdot(P_1+l)}\frac{P_1\cdot P_2}{P_2\cdot l}
H_{5b}^{(0)}\;.
\label{4hai}
\end{eqnarray}

\section*{APPENDIX B: DETERMINATION OF THE PION WAVE FUNCTION}

In this Appendix we comment on the determination of the leading-twist
pion wave function, which can be parametrized as
\begin{equation}
\phi(x)=\frac{3f_\pi}{\sqrt{2N_c}}x(1-x)\left[1+\frac{3}{2}c
\left(5(1-2x)^2-1\right)\right]\;,
\label{pas}
\end{equation}
with the shape parameter $c$. It has been proposed to extract the
leading-twist pion wave function from experimental data of the pion
transition form factor \cite{KR}. The asymptotic model was obtained with
the shape parameter $c\sim 0$,
\begin{equation}
\phi^{AS}(x)=\frac{3f_\pi}{\sqrt{2N_c}}x(1-x)\;.
\label{as}
\end{equation}
The pion wave function can also be determined from other processes
involving pions, such as the pion form factor and the $B$ meson decays
$B\to\pi l{\bar\nu}$ and $B\to D\pi$. It has been known that
a large value of $c$ is preferred for explaining the data of the pion
form factor \cite{LS}.

Another quantity that has been considered is the ratio of the branching
ratios of the $B\to D\pi$ dceays \cite{YL,LM},
\begin{equation}
R\equiv \frac{B(B^+\to D^0\pi^+)}{B(B^0\to D^-\pi^+)}\;.
\end{equation}
The charged $B$ meson decay contains both factorizable and
nonfactorizable amplitudes: the $B\to D$ form factor associated with the
external-$W$ emission, the $B\to\pi$ form factor associated with the
internal-$W$ emission, and the nonfactorizable amplitude associated
with the internal-$W$ emission. The neutral $B$ meson decay contains
only the factorizable external-$W$ emission amplitude. The Wilson
coefficient for the factorizable internal-$W$ emission amplitude is small
at scales around the $b$ quark mass. Hence, the difference between the
branching ratios $B(B^+\to D^0\pi^+)$ and $B(B^0\to D^-\pi^+)$ is
attributed to the nonfactorizable internal-$W$ emission amplitude.
To explain the data of $R\sim 1.6$, a larger $c\sim 0.5$ has been
obtained \cite{LL}. This value of $c$ is located between those for
the asymptotic model and for the Chernyak-Zhinitsky (CZ) model
corresponding to $c=2/3$ \cite{CZ},
\begin{equation}
\phi^{\rm CZ}(x)=\frac{5\sqrt{6}f_\pi}{2}x(1-x)(1-2x)^2\;.
\label{cz}
\end{equation}
Note that the coefficient $c=0.44$ (at the factorization scale about 1
GeV) derived from QCD sum rules \cite{PB1} is close to that extracted
from the ratio $R$. However, also note that a flat pion wave function was
concluded in the framework of covariant quark-pion model \cite{ADT} and
of QCD sum rules \cite{BSM,MR2}.

The above results seem not to be well consistent with the universality of
the pion wave function. We emphasize that the infrared structures of the
processes $\pi\gamma^*\to\gamma$ and $\pi\gamma^*\to\pi$ are different at
next-to-leading twist. For example, the collinear divergences associated
with the pseudo-scalar structure $\gamma_5$ are absent in
$\pi\gamma^*\to\gamma$, but exist in $\pi\gamma^*\to\pi$. The
three-parton wave functions contribute to both the pion transition form
factor and the pion form factor. It is expected that at the maximal energy
scales around 8 GeV$^2$, where data are available, these subleading
contributions are sizeable relative to the leading ones. It has been
explicitly demonstrated that if higher-twist controbutions from parton
transverse momenta are taken into account, the CZ wave function is not
excluded by the data of the pion transition form factor \cite{CHM},
contrary to the conclusion in \cite{KR}. It has been shown that the pion
form factor suffers substantial higher-twist contributions proportional
to the chiral condensate at currently available energy scales \cite{CDH}.
The $B\to\pi$ transition form factor also receives nonvanishing
higher-twsit contributions \cite{KLS}. Because of these
next-to-leading-twist ambiguity, we argue that the above different
extractions of the leading-twist pion wave function should not be
regarded as an inconsistency.

\newpage

{\bf \Large Figure Captions}
\vspace{10mm}

\begin{enumerate}

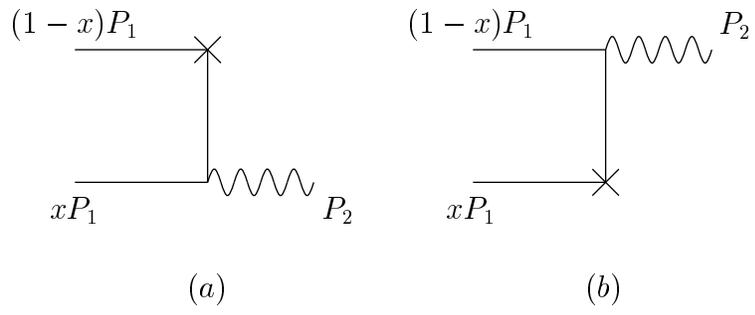
\item Fig. 1: Lowest-order diagrams for $\pi\gamma^*\to\gamma$
($B\to\gamma l\bar\nu$), where the symbol $\times$ represents the
virtual photon (weak decay) vertex.

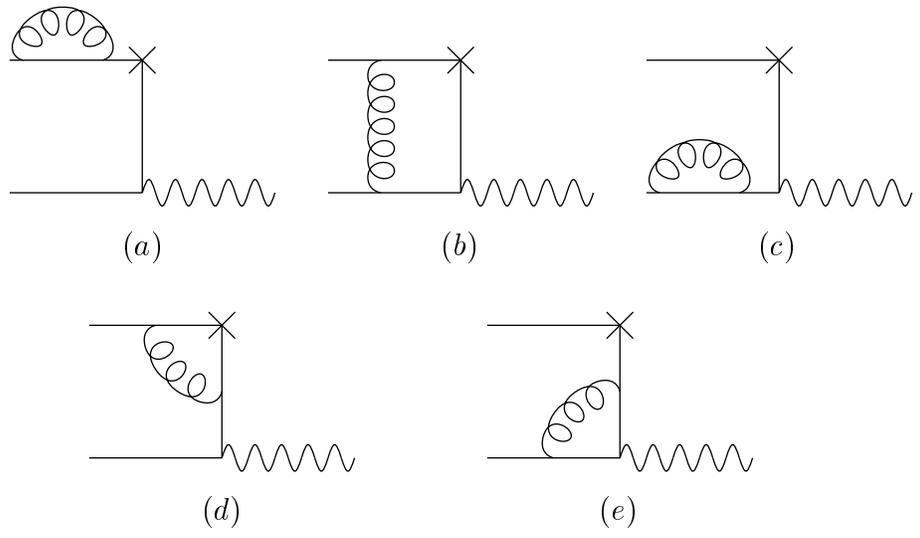
\item Fig. 2: $O(\alpha_s)$ radiative corrections to Fig.~1(a).

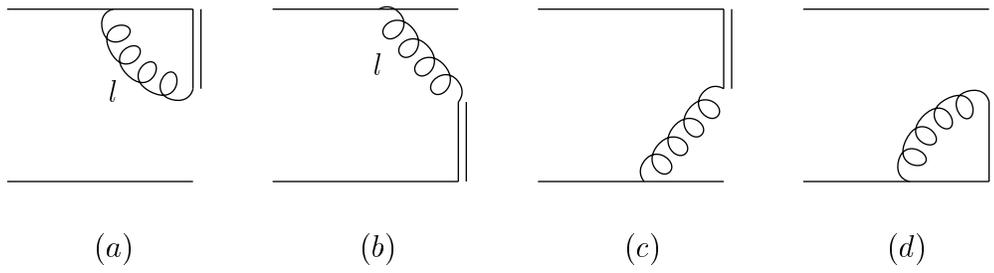
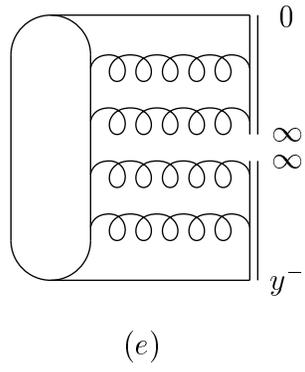
\item Fig. 3: (a)-(d) Infrared divergent diagrams factored out of
Fig.~2(d) and 2(e). (e) The graphic definition of the leading-twist pion
wave function.

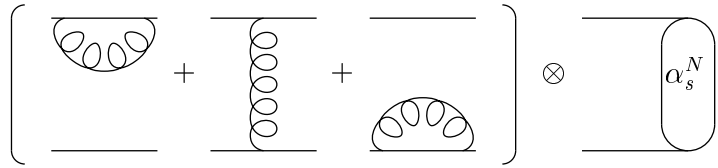
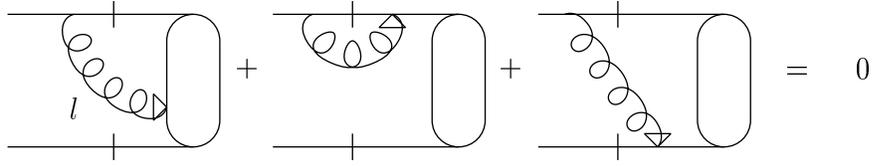
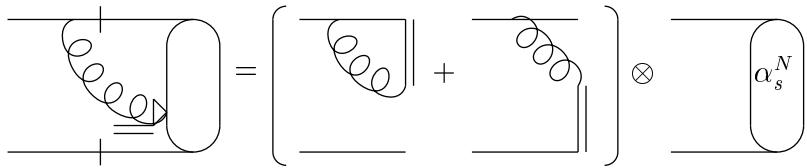
\item Fig. 4: (a) Factorization of $O(\alpha_s^{N+1})$ reducible
diagrams. (b) The Ward identity. (c) Factorization of
$O(\alpha_s^{N+1})$ irreducible diagrams.

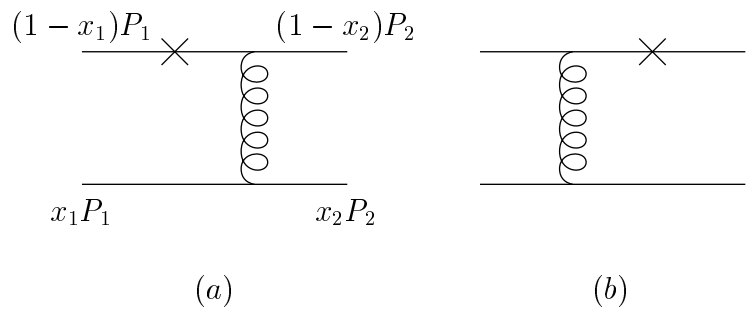
\item Fig. 5: Lowest-order diagrams for $\pi\gamma^*\to\pi$
($B\to\pi l\bar\nu$), where the symbol $\times$ represents the
virtual photon (weak decay) vertex.

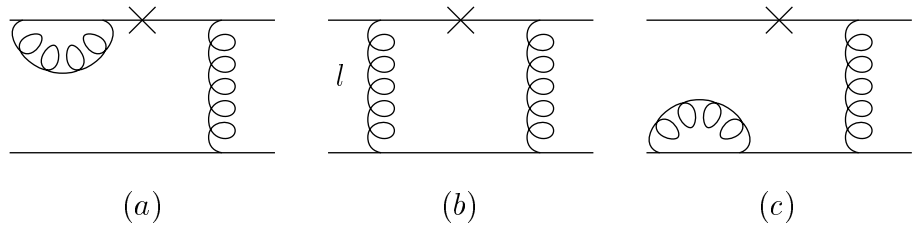
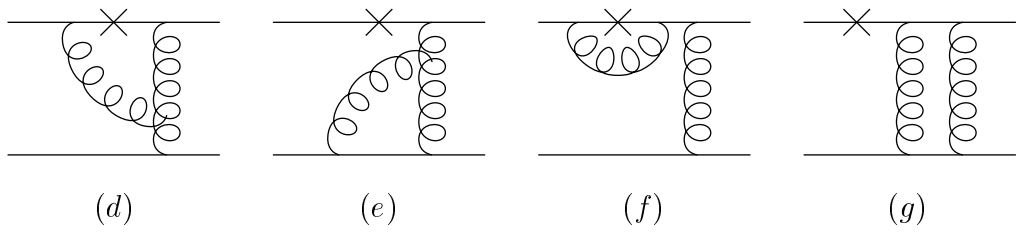
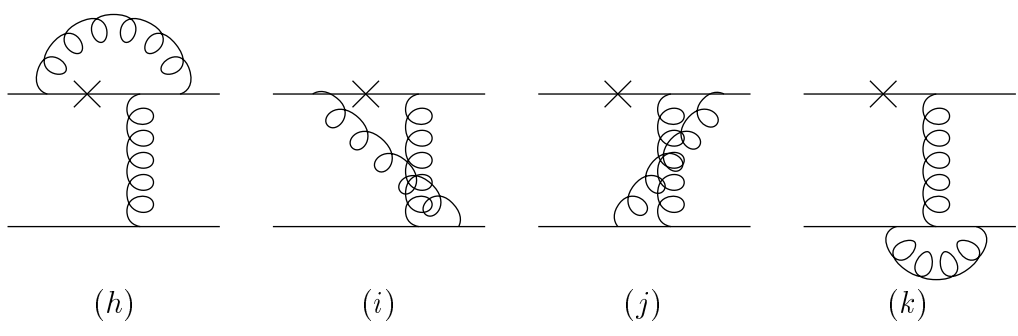
\item Fig. 6: $O(\alpha_s)$ radiative corrections to Fig.~5(a).

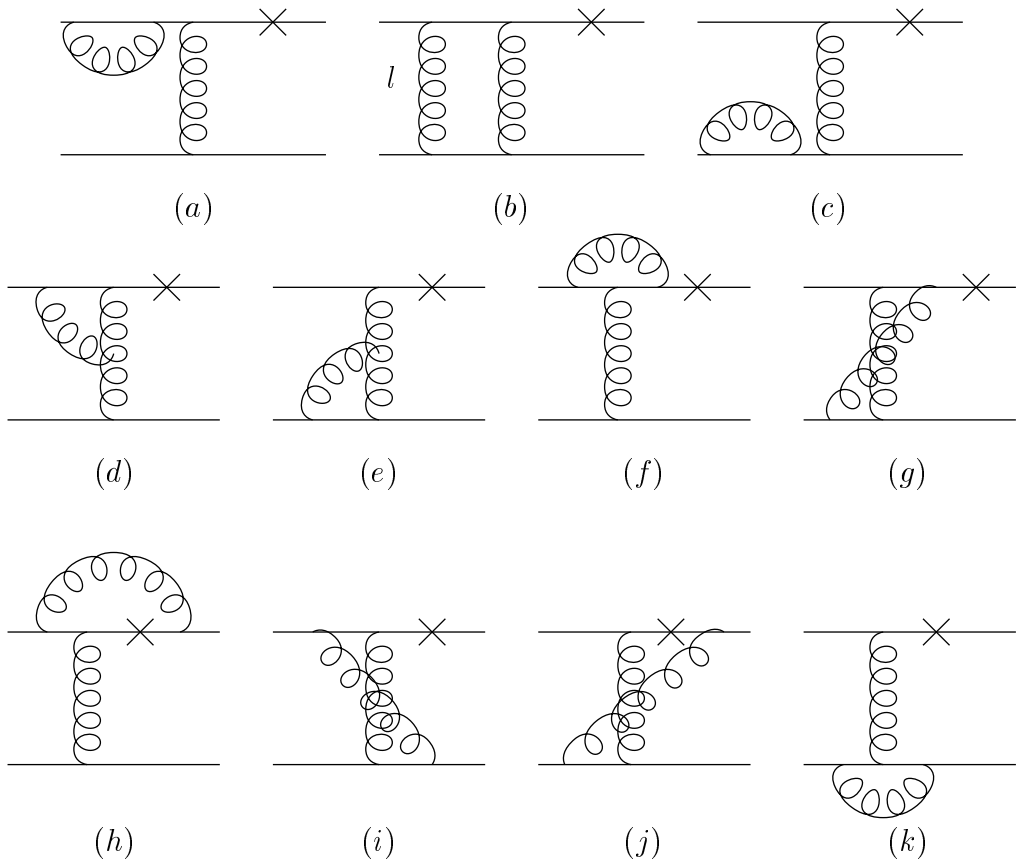
\item Fig. 7: $O(\alpha_s)$ radiative corrections to Fig.~5(b).

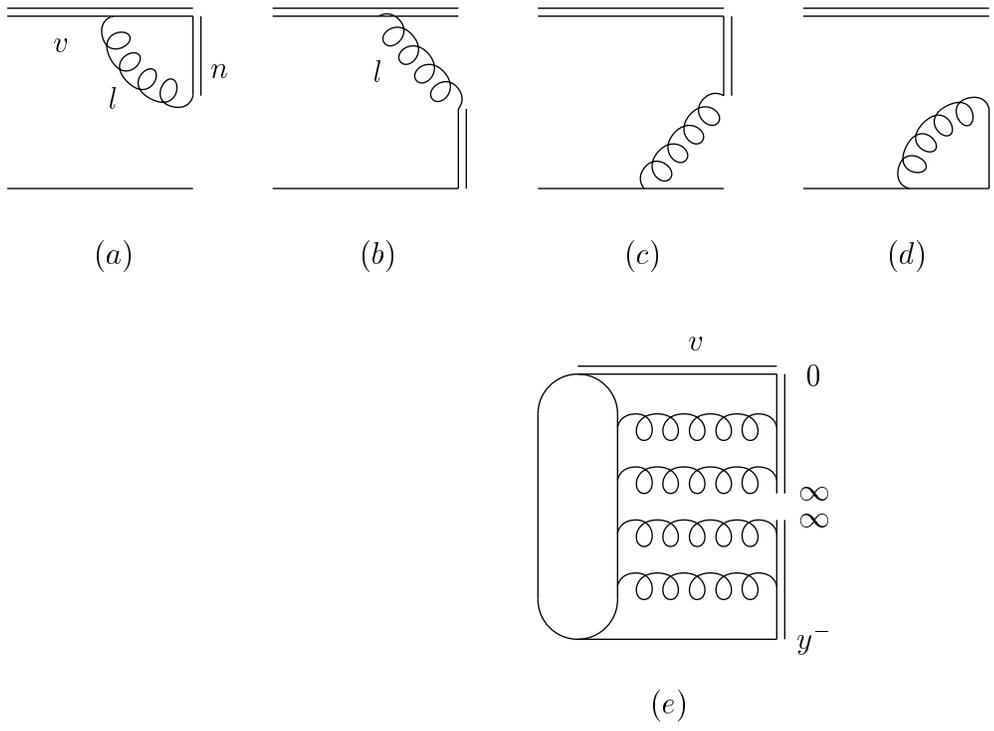
\item Fig. 8: (a)-(d) Infrared divergent diagrams factored out of
Fig.~2(d) and 2(e) for the $B\to\gamma l\bar\nu$ decay. (e) The graphic
definition of the leading-twist $B$ meson wave function.

\end{enumerate}

\end{document}